\RequirePackage{fix-cm}
\documentclass[twocolumn,pdftex,epjc3]{svjour3}

\smartqed  

\usepackage[T1]{fontenc}
\usepackage{fix-cm}
\usepackage{lmodern}
\usepackage{grffile}
\usepackage{textcomp}   
\usepackage{gensymb}
\usepackage[english]{babel}
\usepackage{graphicx}
\usepackage{cuted}
\usepackage{amsmath}
\usepackage{dsfont}\let\mathbb\mathds
\usepackage[breaklinks=true,colorlinks=false,hidelinks]{hyperref}
\usepackage[section]{placeins} 
\usepackage{needspace}
\usepackage{setspace}
\usepackage{tabularx}
\usepackage{subcaption}
\usepackage{multirow}
\usepackage{MnSymbol}
\usepackage{xspace}
\usepackage{pbox}
\usepackage[shortlabels]{enumitem} 
\usepackage{booktabs} 
\usepackage{bigints} 
\usepackage{sidecap} 
\usepackage{makecell} 
\usepackage{csquotes}

\usepackage{tikz}
\usetikzlibrary{arrows.meta,decorations.markings,calc}

\usepackage{physics}
\usepackage[separate-uncertainty = true]{siunitx}

\usepackage{pgfplots}
\pgfplotsset{compat=1.18}

\usepackage{pgfcache}

\usepackage{epstopdf}

\AtBeginDocument{\RenewCommandCopy\qty\SI}

\newcommand{\diff}[1]{\operatorname{d}\ifthenelse{\equal{#1}{}}{\,}{\!#1}}

\newcommand{\MeVc}{\ensuremath{\mathrm{MeV}\!/\!{\it c}}}

\newcommand{\E}{\ensuremath{\mathrm{E}}}

\newcommand{\ecm}{\ensuremath{\si{\elementarycharge}\!\cdot\!\cm}}

\newcommand{\cm}{\ensuremath{\mathrm{cm}}}
\newcommand{\mm}{\ensuremath{\mathrm{mm}}}

\newcommand{\MDM}{{\mathchoice{}{}{\scriptscriptstyle}{}\text{AMM}}}
\newcommand{\EDM}{{\mathchoice{}{}{\scriptscriptstyle}{}\text{EDM}}}
\newcommand{\WF}{{\mathchoice{}{}{\scriptscriptstyle}{}\text{WF}}}
\newcommand{\Kick}{{\mathchoice{}{}{\scriptscriptstyle}{}\mathrm{K}}}
\newcommand{\CW}{{\mathchoice{}{}{\scriptscriptstyle}{}\mathrm{CW}}}
\newcommand{\CCW}{{\mathchoice{}{}{\scriptscriptstyle}{}\mathrm{CCW}}}

\newcommand{\peak}{{\mathchoice{}{}{\scriptscriptstyle}{}}}

\journalname{Eur. Phys. J. C}

\begin{document}

\title{Anomalous spin precession systematic effects in the search for a muon EDM using the frozen-spin technique}

\author{
	G. Cavoto\thanksref{addr3}
	\and
	R. Chakraborty\thanksref{addr1}
	\and
	A. Doinaki\thanksref{addr1,addr5}
	\and
	C. Dutsov\thanksref{addr1, email}
	\and
	M. Giovannozzi\thanksref{addr2}
	\and
	T. Hume\thanksref{addr1,addr5}
	\and
	K. Kirch\thanksref{addr1,addr5}
	\and
	K. Michielsen\thanksref{addr1,addr5,ecpo}
	\and
	L. Morvaj\thanksref{addr1}
	\and
	A. Papa\thanksref{addr4}
	\and
	F. Renga\thanksref{addr3}
	\and
	M. Sakurai\thanksref{addr5, addr6}
	\and
	P. Schmidt-Wellenburg\thanksref{addr1}
}

\thankstext{email}{e-mail: chavdar.dutsov@psi.ch}

\institute{%
	Paul Scherrer Institut, CH-5232 Villigen PSI, Switzerland \label{addr1}
	\and
	CERN Beams Department, Esplanade des Particules 1, 1211 Meyrin, Switzerland \label{addr2}
	\and
	Istituto Nazionale di Fisica Nucleare, Sez. di Roma, P.le A. Moro 2, 00185 Rome, Italy \label{addr3}
	\and
	Istituto Nazionale di Fisica Nucleare, Sez. di Pisa, Largo B. Pontecorvo 3, 56127 Pisa, Italy \label{addr4}
	\and
	ETH Z\"urich, 8092 Z\"urich, Switzerland \label{addr5}
	\and
	\'Ecole Polytechnique, Route de Saclay, 91128 Palaiseau Cedex, France\label{ecpo}
	\and
	\emph{Present Address:} University College London, Gower Street, London, WC1E 6BT, United Kingdom \label{addr6}
}

\date{Received: date / Accepted: date}

\maketitle

\begin{abstract}
	At the Paul Scherrer Institut~(PSI), we are currently working on the development of a high-preci\-sion apparatus with the aim of searching for the muon electric dipole moment~(EDM) with unprecedented sensitivity. The underpinning principle of this experiment is the frozen-spin technique, a method that suppresses the spin precession due to the anomalous magnetic moment, thereby enhancing the signal-to-noise ratio for EDM signals. This increased sensitivity facilitates measurements that would be difficult to achieve with conventional $g - 2$ muon storage rings. Given the availability of the $p = \SI{125}{MeV/\textit{c}}$ muon beam at PSI, the anticipated statistical sensitivity for the EDM after a year of data collection is \SI{6e-23}{\ecm}. To achieve this goal, it is imperative to meticulously analyse and mitigate any potential spurious effects that could mimic EDM signals. In this study, we present a quantitative methodology to evaluate the systematic effects that might arise in the context of employing the frozen-spin technique within a compact storage ring. Our approach entails the analytical derivation of equations governing the motion of the muon spin in the electromagnetic~(EM) fields intrinsic to the experimental setup, validated through subsequent numerical simulations. We also illustrate a method to calculate the cumulative geometric (Berry's) phase. This work complements ongoing experimental efforts to detect a muon EDM at PSI and contributes to a broader understanding of spin-precession systematic effects.
\end{abstract}


\section{Introduction}

The existence of a permanent EDM in any elementary particle suggests a violation of Charge-Parity~(CP) symmetry.
Within the framework of the Standard Model (SM) of particle physics, EDMs are predicted to be remarkably small, despite the substantial CP-violating phase provided by the Cabibbo-Kobayashi-Maskawa matrix.
In fact, they are so small that they are beyond the reach of any imminent measurements.
Nevertheless, numerous SM extensions allow for substantial CP violating phases, which can result in large EDMs~\cite{Pospelov:2013sca,Seng:2014lea}.
Recently, the EDM of the muon has drawn significant attention, due to a
persistent tension between the experimental results for the muon anomalous
magnetic moment (AMM)~\cite{Abi2021PRL,FNAL2023PRL} and the theoretical SM predictions~\cite{Aoyama:2020ynm}.

Farley et al.~\cite{Farley2004PRL,Khriplovich1998,Semertzidis2001} proposed a method to measure EDMs in storage rings, known as the frozen-spin technique.
The frozen-spin technique cancels the anomalous $(g-2)$ precession by applying a radial electric field perpendicular to the momentum of the stored particles and to the magnetic field, so that any remaining precession is a consequence of the EDM\@.
In a real-world storage ring, where precession due to the AMM cannot be completely suppressed, EDM-like signals may be induced. Such systematic effects can reduce experimental sensitivity or result in a signal mimicking a genuine EDM\@.

A non-zero EDM manifests itself through a precession of the spin around the electric-field vector in the particle's frame of reference.
In the case of muons, the spin precession can be measured by studying the direction of the emitted decay positrons, which is correlated to the spin direction.
%
Our study is focused on the systematic effects induced by coupling of the magnetic dipole moment to the EM field of the experiment.
We delve into both the dynamic and geometric phases of the spin of muons circulating within the confines of a compact storage ring that employs the frozen-spin technique.
Although we strive to maintain sufficient generality in our derivations to ensure their applicability across different scenarios, our discussions are rooted in the context of the ongoing experimental efforts.
Specifically, we have evaluated potential systematic effects associated with the
ongoing effort to search for a muon EDM at PSI\@.

The main part of this paper is separated in four sections. First we present the
details of the experimental setup which will be used to search for the EDM of
the muon. The frozen-spin technique is elaborated upon and the expected
statistical sensitivity is given. In the next section we derive analytical
expressions for the motion of the muon spin in an idealised version of the
EM fields of the experiment. The results are verified by comparison with simulations using \texttt{Geant4}~\cite{Agostinelli2003,Allison2006,Allison2016}, shown in detail in the appendix.
The derivations treat both the dynamic phase build-up, as well as the potential for generation of geometric phases.
Next we consider effects arising from possible
deviations of the real EM field from the nominal one.
In this part we also derive effects arising from deviations of the initial spin and momentum of the muon at the start of a measurement.
Finally, in the discussion, we use the derived analytical
equations to calculate limits on parameters of the experimental setup such that
any possible systematic effect is lower than our target sensitivity for the search for a
muon EDM\@.

\section{Search for the muon EDM at PSI}

The search for a muon EDM at PSI will rely on a storage ring
inside a compact solenoid with inner diameter less than a meter~\cite{Adelmann2010JPG,Adelmann2021arXiv}. The muons will be injected
into the solenoid one by one, through a superconducting injection
channel~\cite{Barna2017} and subsequently kicked by a pulsed magnetic field into a
stable orbit within a weakly-focusing
field~\cite{Iinuma2016}. Two concentric cylindrical electrodes will provide a radial electric field at the position of the muon orbit. The strength of this electric field must be precisely tuned so as to satisfy the frozen-spin condition, where the anomalous spin precession is cancelled and the spin remains aligned with the muon momentum for the duration of its lifetime.

\begin{figure*}[ht]
	\centering
	\includegraphics[width=\linewidth]{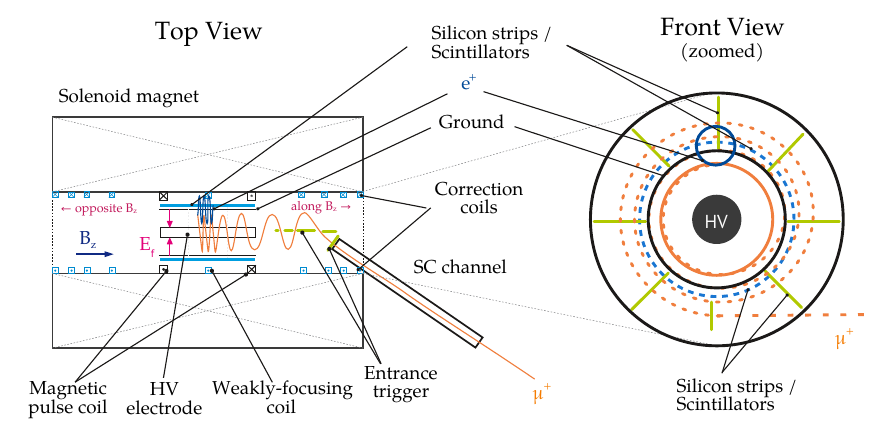}
	\caption{Illustration of the Phase~I muon EDM experimental device employing a compact storage ring inside a solenoid magnet.}
	\label{fig:phaseIexperiment}
\end{figure*}

As the muon decays, the direction of its spin can be statistically inferred from
the trajectory of the emitted decay positron. The parity violation in the weak decay results in a preference for high-energy positrons to be emitted in the direction
of the muon spin. The EDM will be calculated based on the change in asymmetry,
$dA/dt$, where $A(t) = (N_\uparrow(t) - N_\downarrow(t))/(N_\uparrow(t) + N_\downarrow(t))$, which measures the
difference between the number of positrons emitted along or opposite the main magnetic field, see
Fig.~\ref{fig:phaseIexperiment}. Detectors positioned symmetrically on both
sides of the plane defined by the ideal muon orbit will monitor the direction of
emission.

A staged approach has been adopted for this project. The initial phase
(Phase~I) will focus on demonstrating the feasibility of all critical
techniques, with the goal of achieving a sensitivity better than $\sigma(d_\mu)\leq\SI{3e-21}{\ecm}$ to the muon EDM\@. The next phase (Phase~II) aims to achieve a sensitivity of better
than \SI{6e-23}{\ecm}, which would represent an improvement of more than
three orders of magnitude over the current experimental limit of $d_\mu\leq\SI{1.8e-19}{\ecm}$~(C.L. 95\%)~\cite{Bennett2009PRD}.

To reach this target sensitivity, it is crucial to ensure that potential
systematic effects leading to a false EDM signal are adequately controlled. In
particular, we investigate the impact of EM field irregularities on the
experimental results. For this, we study the relativistic spin motion of a positively charged ($+e$) muon of mass $m$ with momentum $\vec{p}$ in electric
$\vec{E}$ and magnetic $\vec{B}$ fields described by the Thomas-BMT
equation~\cite{Thomas1926,Thomas1927,PhysRevLett.2.435}, with an additional term
describing the effect of the EDM, namely
\begin{equation}\label{eq:omegaMuWithEDM}
	\begin{split}
		\vec{\Omega} = &~ \vec{\Omega}^\MDM + \vec{\Omega}^\EDM =\\
		= &-\frac{e}{m}\left[a\vec{B}-\frac{a\gamma}{\left(\gamma+1\right)}\left(\vec{\beta}\cdot\vec{B}\right)\vec{\beta}-\left(a+\frac{1}{1-\gamma^2}\right) \frac{\vec{\beta}\times\vec{E}}{c}\right] \\
		& -\frac{\eta e}{2m}\left[\vec{\beta}\times\vec{B}+\frac{\vec{E}}{c}-\frac{\gamma}{c(\gamma+1)}\left(\vec{\beta}\cdot\vec{E}\right)\vec{\beta}\right]\,,
	\end{split}
\end{equation}
where $\vec{\beta}=\vec{p}c/E$ and $\gamma=\left(1-\beta^2\right)^{-1/2}$ are the relativistic factors with total energy $E$, $a$ the anomalous magnetic moment, and $\eta=4d_\mu m c/(e\hbar)$ the gyro-electric ratio multiplied by $2mc/e$ and is the dimensionless constant describing the size of the EDM.

The second line of Eq.~\eqref{eq:omegaMuWithEDM} represents the anomalous
precession frequency $\vec \Omega^\MDM$, the difference between the Larmor
precession and the cyclotron precession, due to the AMM\@. The last line represents the precession $\vec\Omega^{\EDM}$ due to the EDM coupling to the electric field in the boosted reference frame of the moving muon.

The experimental setup proposed for the search for a muon EDM is based on the
ideas and concepts discussed in
\cite{Farley2004PRL,Adelmann2010JPG,Adelmann2021arXiv,Iinuma2016}. The
salient feature of the proposed search is the cancellation of the precession due
to the anomalous magnetic moment by meticulously choosing a radial electric
field, and thus fully exploiting the large electric field $\gamma c\vec{\beta}\times \vec{B}\approx \SI{1}{GV/m}$ in the rest frame of the muon to achieve a perpendicular precession ($\vec{\Omega} \, \bot \,	\vec{B}$) only. By examining Eq.~\eqref{eq:omegaMuWithEDM}, we can counteract the anomalous precession term by applying an electric field such that:
\begin{equation}
	a\vec{B} = \left(a+\frac{1}{1-\gamma^2}\right)\frac{\vec{\beta}\times\vec{E_\mathrm{f}}}{c}.
	\label{eq:FrozenSpinCondition}
\end{equation}
In the case of $\vec{\beta}\cdot\vec{B}=\vec{\beta}\cdot\vec{E}=0$, $\vec{B}\cdot\vec{E}=0$, and assuming $a \ll 1/(1-\gamma^2)$, which is a good approximation for small $\gamma$,  we find a required field strength of  $\vert E_{\rm f} \vert \approx a c\beta\gamma^2 \vert B \vert$.  Hence, by selecting the exact field condition of
Eq.~\eqref{eq:FrozenSpinCondition}, the cyclotron precession frequency is modified such that the relative angle between the momentum vector and the spin
remains unchanged if $\eta =0$; the spin is ``frozen''.

\subsection{Sensitivity to the muon EDM}

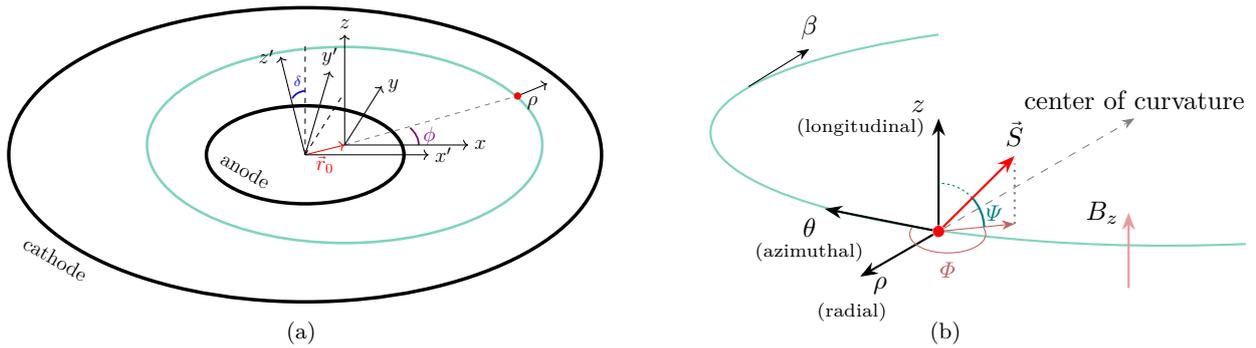
\begin{figure*}[ht]
	\centering
	\subfloat[]{
		\scalebox{0.65}{
			\begin{tikzpicture}[
					vector/.style={-Stealth, thick, black},
					vector2/.style={-Stealth, thin, dashed, gray},
					particle/.style={circle, fill=red, inner sep=1.5pt},
					vector_label/.style={black, font=\footnotesize},
					large_label/.style={font=\large},
					orbit/.style={thick, aqua}
					scale=1
				]
				\definecolor{aqua}{rgb}{0.5, 0.83, 0.75}

				\coordinate (dispCoord) at (-0.8,-0.2);

				\draw[line width=2pt] (dispCoord) ellipse (6cm and 3cm);
				\node[large_label, anchor=north east, rotate=-30] at (-5,-2.5) {cathode};
				\draw[line width=1.5pt, aqua] (0,0) ellipse (4cm and 2cm);
				\draw[line width=2pt] (dispCoord) ellipse (2cm and 1cm);
				\node[large_label, anchor=north east, rotate=-30] at (-1.3,-0.6) {anode};
				\draw[->,line width=0.6pt] (3.5,1) -- (4.1,1.25) node[large_label,midway, below] {$\rho$};
				\draw[dashed,line width=0.2pt, gray] (0,0) -- (3.5,1) node[midway, below] {};
				\draw[violet, line width=1pt] (1.5,0) arc (0:28:1.5 and 0.75)
				node[large_label, midway, right, text=violet] {$\phi$};

				\node[particle] (P) at (3.5,1) {};

				\draw[->,line width=0.6pt] (0,0) -- (2.5,0) node[large_label, anchor=west] {$x$};
				\draw[->,line width=0.6pt] (0,0) -- (0.75,1.2) node[large_label, anchor=west] {$y$};
				\draw[->,line width=0.6pt] (0,0) -- (0,2.25) node[large_label, anchor=south] {$z$};

				\draw[->, line width=0.6pt, red] (dispCoord) -- (0,0) node[large_label, midway, below] {$\vec r_0$};
				\draw[->,line width=0.6pt] (dispCoord) -- ++(2.5,0) node[large_label, anchor=west] {$x'$};
				\draw[->,line width=0.6pt] (dispCoord) -- ++(0.5,1.7) node[large_label, anchor=south] {$y'$};
				\draw[dashed,line width=0.6pt] (dispCoord) -- ++(0.75,1.2) node[large_label, anchor=west] {};
				\draw[dashed,line width=0.6pt] (dispCoord) -- ++(0,2.25) node[large_label, anchor=south] {};
				\draw[->,line width=0.6pt] (dispCoord) -- ++(-0.5,2.) node[large_label, anchor=east] {$z'$};
				\draw[blue, line width=1pt] ($(dispCoord) + (-0, 1.3)$) arc (90:150:0.3 and 0.3) node[vector_label, midway, above, text=blue] {$\delta$};
			\end{tikzpicture}
		}
		\label{fig:coord_system_a}
	}
	\subfloat[]{
		\vspace*{-10pt}
		\scalebox{1.}{
			\begin{tikzpicture}[
					vector/.style={-Stealth, thick, black},
					vector2/.style={-Stealth, thin, dashed, gray},
					particle/.style={circle, fill=red, inner sep=1.5pt},
					vector_label/.style={black},
					orbit/.style={thick, aqua}
				]

				\definecolor{salmon}{rgb}{0.918, 0.6, 0.6}
				\definecolor{darkPastelRed}{rgb}{0.76, 0.38, 0.38}
				\definecolor{aqua}{rgb}{0.5, 0.83, 0.75}

				\draw[orbit] (0,0) arc (120:280:6 and 1.5);

				\node[particle] (P) at (0,-2.61) {};

				\draw[vector2] (P) -- ++(2.58,1.5) node[vector_label, above] {center
					of curvature};
				\draw[vector] (P) -- ++(-1.5,+0.3) node[vector_label, below, xshift=-.2cm, align=center] {$\theta$\\[-0.5em] \scriptsize{(azimuthal)}};
				\draw[vector] (P) -- ++(-1.032,-0.6) node[vector_label, right, align=right, xshift=-0.7cm, yshift=-0.3cm] {$\rho$\\[-0.2em] \scriptsize{(radial)}};
				\draw[vector] (P) -- ++(0,1.5) node[vector_label, right, align=right, xshift=-2cm] {$z$\\[-0.5em] \scriptsize{(longitudinal)}};

				\draw[vector, salmon] ($(P)+(+2.5,-0.75)$) -- ++(0,1) node[vector_label, left]
				{$B_z$};
				\draw[vector, black, thin] ($(P)+(-2.5,1.92)$) -- ++(0.8,0.5) node[vector_label,
					above]
				{$\beta$};

				\draw[vector, darkPastelRed, thin] (P) -- ++(1,0.1);
				\draw[vector, gray, dotted, -] ($(P) + (1, 0.1)$) -- ++(0.0,0.8);

				\draw[darkPastelRed] ($(P) + (-0.35, 0.05)$) arc (-200:20:0.5 and 0.25)
				node[vector_label, midway, below, text=darkPastelRed]
					{\footnotesize$\Phi$};
				\draw[teal, thin, dash pattern={on 1pt off 1pt}] ($(P) + (0.6, 0.06)$) arc (2:90:0.6 and
				0.6);
				\draw[teal, thick] ($(P) + (0.6, 0.06)$) arc (2:43:0.6 and
				0.6)
				node[vector_label, midway, text=teal, xshift=0.18cm,
						yshift=-0.02cm]
					{\footnotesize$\Psi$};
				\draw[vector, red, thick] (P) -- ++(1,1) node[vector_label, above] {$\vec S$};

			\end{tikzpicture}
		}
		\label{fig:coord_system_b}
	}
	\caption[Local reference coordinate system.]{Representations of the
		reference frames used throughout the paper. a) Cartesian reference frame used to describe the $E$-field in
		the particle rest frame. The origin of the $(x, y, z)$ reference
		frame coincides with the centre of a given muon orbit, where the $z$
		axis is parallel to the field of the main solenoid and the $xy$
		plane lies in the orbit plane. The origin of the primed reference
		frame is at the centre of the cylindrical electrodes, where $z'$
		runs parallel to the central axes of the inner and outer electrodes
		(anode and cathode). The
		angle $\delta$ is the angle between the central axes of the
		electrodes and the main solenoid.
		b) The curvilinear (Frenet-Serret) reference coordinate system used to derive the
		motion of the spin in the EM fields of the experiment. The axis
		$\theta$ follows the momentum of the muon and $z$ is always parallel to the
		main solenoid magnetic field $B_z$
		The
		vector $\vec S$ is the normalised particle spin. The angle $\Phi$ is the azimuthal
		spin phase (in the plane of the orbit), and $\Psi$ is the complementary angle to the polar angle.
	}
	\label{fig:coord_system}
\end{figure*}

Using Eq.~\eqref{eq:omegaMuWithEDM} and assuming that $\vec \beta \cdot
	\vec E = 0$, $\vec \beta \cdot \vec B = 0$ and $\lvert E \rvert \ll c\lvert
	\vec\beta\times \vec B\rvert$ (as evident from the aforementioned
\SI{0.3}{MV/m} $\ll$ \SI{1}{GV/m}), the spin precession angular velocity due to a non-zero EDM is:
\begin{equation}
	\vec \Omega^\EDM = \frac{\eta}{2}\frac{e}{m}\vec\beta \times \vec B,
	\label{eq:edm_angular_velocity}
\end{equation}
Note that the coordinate system used here and
throughout this work is such that it follows the reference particle orbit
(similar to~\cite{HajTahar2021,Rabi1954}) as sketched in
Fig.~\ref{fig:coord_system_b}.
The initial orientation of the spin $\vec S = (S_\theta, S_\rho, S_z)$ in spherical
coordinates is
\begin{equation}
	\Phi_0   = \arctan\left( \frac{S_\rho}{S_\theta} \right), \quad
	\Psi_0 = \frac{\pi}{2} - \arccos\left( S_z \right),
	\label{eq:phi_theta_initial}
\end{equation}
where $\lvert \vec S \vert = 1$, $\Phi$ is the azimuthal spin phase, i.e., in the plane of the orbit, and $\Psi$ is the complementary angle to the polar angle.

From the assumption that the $E$-field and the $B$-field are perpendicular to each other and to the muon velocity, the only sizeable component of $\vec\Omega_\EDM$ is the radial component,
\begin{equation}
	\Omega^\EDM_\rho = \dot \Psi = \frac{2c}{\hslash}\beta_\theta B_z d_\mu,
	\label{eq:edm_angle_vs_de}
\end{equation}
where we have replaced $\eta$ with the expression for $d_\mu$.

In Phase~I and Phase~II we will store muons with $\beta_\theta = 0.256$ ($p =
	\SI{28}{MeV/\textit{c}}$) and $0.764$ ($p = \SI{125}{MeV/\textit{c}}$),
respectively, in a magnetic field of strength $B_z=\SI{3}{T}$.
This results in angular velocities for an EDM equal to the statistical sensitivity of:
\begin{align}
	\dot \Psi_\mathrm{I}   = \SI{21.15}{ rad/s} & \text{\quad for\quad}  d_\mu =
	\SI{3e-21}{\ecm},
	\label{eq:limit_on_angular_velocity_prec}                                                        \\
	\dot \Psi_\mathrm{II}  = \SI{1.26}{ rad/s}  & \text{\quad for\quad }  d_\mu  = \SI{6e-23}{\ecm}.
	\label{eq:limit_on_angular_velocity_final}
\end{align}
The radius of the orbit is $\rho_0 = \SI{31}{mm}$ for the first phase and \SI{134}{mm} for the second. The required frozen spin field is $E_\mathrm{f} = \SI{287}{kV/m}$ and $\E_\mathrm{f} = \SI{1.92}{MV/m}$ for Phase~I and II respectively.


\section{The spin motion of muons in the experiment}
As a starting point to the analysis of possible systematic effects we derive an approximate analytical expression for the spin motion in the field configuration characteristic of the frozen-spin technique.
For this purpose we approximate the magnetic field of the solenoid in the region of the storage ring as a uniform magnetic field oriented along the $z$-axis; the weakly-focusing field by the first-order approximation of a field generated
by a circular coil; and, the electric field as a radial field produced by the potential difference between two infinite coaxial cylindrical
electrodes.
We then parameterise the most important and most likely imperfections of these fields and estimate their effect on the spin precession in the following sections.

\subsection{Spin precession around the radial axis}\label{sec:weakly_focusing_desc}

The longitudinal position of a particle with charge $e$, mass $m$, and velocity
$c\vec\beta$ is given by the solution of:
\begin{equation}
	\ddot{z} = \frac{e}{\gamma m} \left(E_z + c \beta_\theta B_\rho(z) +
	c\beta_\rho B_\theta(z) \right),
	\label{eq:lorentz}
\end{equation}
where $B_\rho(z) \approx z \frac{\partial B_\rho(\rho_0)}{\partial z} = z \partial_z
	B_\rho(\rho_0)$, and we assume that a constant non-zero $z$-component of the
electric field exists. In general, the last term $\beta_\rho B_\theta(z) \ll
	\beta_\theta B_\rho(z)$ as $\beta_\rho$ is practically zero for stored
particles and $B_\theta$ is zero if there is no electrical current flowing
through the area enclosed by the orbit. Therefore, the term is ignored in
the further discussion and the solution of the differential equation becomes that
of a harmonic oscillator with longitudinal displacement
\begin{equation}
	z(t) = z_0\cos (\omega_\mathrm{b} t + \varphi_0) +
	\frac{1}{\partial_z B_\rho}\frac{E_z}{c\beta_\theta},
	\label{eq:omega_betatron_calc}
\end{equation}
where $\omega_\mathrm{b}$ is the angular velocity of the longitudinal betatron oscillation.
It can be expressed in terms of the field gradient index
$n = \frac{\rho_0}{B_0}\partial_z B_\rho$ as $\omega_\mathrm{b} = \omega_\mathrm{c} \sqrt{n}$,
where $\omega_\mathrm{c}= -eB_0/\gamma m$ is the cyclotron angular velocity, $\rho_0$ is
the radius of the nominal orbit, and $B_0$ is the magnetic field of the main
solenoid. The particles in the storage ring also experience a horizontal
betatron oscillation (in the plane of the orbit), with angular velocity
$\omega_\mathrm{h} = \omega_\mathrm{c} \sqrt{1 - n}$, that corresponds to
oscillations in $B_z$ and does not directly lead to spin precession
mimicking the EDM signal. In a compact storage ring configuration $\omega_\mathrm{h} \approx \omega_\mathrm{c}$ since $n \ll 1$, such that a small difference
between these two frequencies leads to a slow precession of the muon orbit whose effects
are explored in section~\ref{sec:longitudinal_efield}.


The relative precession of the spin due to the coupling of the AMM to the radial
magnetic field of the weakly-focusing field is
\begin{equation}
	\begin{split}
		\Omega^\WF_\rho &= -\frac{ea}{m}B_\rho(z(t)) \approx \\
		&\approx	-\frac{ea}{m}\left[\partial_z B_\rho  z_0 \cos(\omega_\mathrm{b} t + \varphi_0) - \frac{1}{c\beta_\theta} E_z\right],
		\label{eq:precession_weakly_focusing}
	\end{split}
\end{equation}
where the index $\rho$ denotes the radial component of $\vec \Omega$.

Another source of radial precession that has to be considered is the radial
magnetic field in the reference frame of the muon due to a non-zero
longitudinal electric field in the laboratory reference frame. For a non-zero longitudinal electric field, $|E_z|>0$, we obtain
\begin{equation}
	\Omega^{\scriptscriptstyle{}E_z}_\rho
	=
	-\frac{e}{mc} \left( a - \frac{1}{\gamma^2 -1} \right)
	\beta_\theta E_z,
	\label{eq:precession_ey}
\end{equation}
for the radial component only, by applying the T-BMT equation.

Further, a radial spin precession could also be caused by a radial $B$-field $B^\Kick_\rho$ caused by residual currents in coils or eddy currents induced by the short, $\Delta t_{\rm pulse}\approx\SI{100}{ns}$, magnetic pulse used to kick muons onto a stable orbit, see Fig.~\ref{fig:phaseIexperiment}.
This field can be described by a superposition of periodic oscillations,
\begin{equation}
	B^{\Kick}_\rho(t) = \int_0^\infty A_\rho(\omega) \cos(\omega t + b_0(\omega)) d\omega,
	\label{eq:b_rho_arbitrary}
\end{equation}
where $A_\rho(\omega)$ is the oscillation amplitude as a function of the angular frequency $\omega$ and $b_0(\omega)$ is an arbitrary frequency-dependent phase.
Possible systematic effects due to such oscillations are explored in detail in section~\ref{sec:uniformities}.

Combining Eqs.~\eqref{eq:precession_weakly_focusing} and~\eqref{eq:precession_ey}, and including the term for an arbitrary radial
magnetic field $B^\Kick_\rho(t)$ as in Eq.~\eqref{eq:b_rho_arbitrary}, one
obtains the total angular velocity of the radial precession due to the AMM
around the $\rho$-axis
\begin{multline}
	\Omega^\MDM_\rho =
	-\frac{ea}{m} \left[
		\frac{1}{c}\left( 1-\frac{1}{a(\gamma^2 - 1)} -
		\frac{1}{\beta^2_\theta}\right) \beta_\theta E_z \right.\\ +
		\partial_z B_\rho z_0 \cos(\omega_\mathrm{b} t + \Phi_0)  + B^\Kick_\rho(t)\bigg].
	\label{eq:precession_mdm_total_rho}
\end{multline}

We are interested in the average angular velocity over many muon orbits. In this
case, the average due to the betatron oscillations is zero, as $\langle \cos
	(\omega_\mathrm{b} t) \rangle = 0$ for $t \gg \omega_\mathrm{b}^{-1}$.
In general, the azimuthal
velocity $\beta_\theta$ and the longitudinal electric field $E_z$ are not
correlated, thus the average over time of their product is the product of their
averages
\begin{multline}
	\left\langle \Omega^{\MDM}_\rho \right\rangle = \\
	-\frac{ea}{mc} \left\langle \left(1 - \frac{1}{a(\gamma^2 - 1)} -
	\frac{1}{\beta_\theta^2}\right) \beta_\theta \right\rangle \langle E_z \rangle -
	\frac{ea}{m}\left\langle B^\Kick_\rho(t) \right\rangle.
	\label{eq:precession_rho_avg}
\end{multline}

Note that here $E_z$ is a static uniform field and $B^\Kick_\rho$ is an
arbitrary time-dependent field. Effects of their time stability and spatial
uniformity are discussed in section~\ref{sec:uniformities}.

Although the average of the betatron oscillations is zero, oscillations can
still occur around two perpendicular axes, potentially leading to the
accumulation of a geometrical, also known as Berry's, phase~\cite{Berry1984}. Additionally, potential
systematic effects may arise from the approximation $t \gg
	\omega_\mathrm{b}^{-1}$.
These sources of systematic effects are explored in more detail in section~\ref{sec:berry_phases}.

\subsection{Azimuthal spin precession}
When the muons circulate in the storage ring, they oscillate longitudinally
around an equilibrium orbit (betatron oscillation). The equilibrium orbit is
perpendicular to the longitudinal magnetic field.
In the absence of other fields the betatron oscillation results from  the weakly-focusing field.
Due to this betatron motion, the momentum vector of the particle is not at all times perpendicular to the longitudinal magnetic field, leading to a non-zero projection of the magnetic field along its trajectory.
This field is proportional to the angle, $\zeta = \pi/2 - \angle (\vec\beta, \vec B)$, between the muon momentum and the plane of the equilibrium orbit. In turn,
$\tan \zeta	= p_z / p_\theta \approx \sin \zeta$ and $p_z$ oscillates as $p_z = p_{z_0} \sin
	(\omega_\mathrm{b} t)$.
The muon is therefore exposed to an oscillating azimuthal field,
\begin{equation}
	B_\theta(t) = -B_z \sin \zeta \approx  -B_z \frac{p_{z_0}}{p_\theta} \sin( \omega_\mathrm{b} t),
	\label{eq:b_theta}
\end{equation}
where the momentum
\begin{equation}
	p_{z_0} = e c \beta \partial_z B_\rho z_0 \int_0^{\frac{\pi}{2 \omega_\mathrm{b}}}
	\cos (\omega_\mathrm{b} t)  dt =  \frac{e c \beta z_0}{\omega_\mathrm{b}} \partial_z B_\rho,
	\label{eq:p_z_0}
\end{equation}
is the $z$-component of the  momentum at $z=0$.
This also means that there
will be a non-zero $z$-component of the velocity, given by
\begin{equation}
	\beta_z(t) = \frac{p_{z_0}}{p_\theta}\,\beta_\theta \sin( \omega_\mathrm{b} t) .
	\label{eq:beta_z}
\end{equation}

If the radial electric field $E_\rho$ is correctly set to the value
$E_\mathrm{f}$ required for the frozen-spin technique, then there will be no
oscillations around the azimuthal $\theta$-axis as the electric field perfectly
counteracts the precession induced by the coupling of the AMM to the longitudinal
field of the solenoid. However, if $E_\rho \neq E_\mathrm{f}$ there will be imperfect
cancellation of the $g-2$ precession around $\theta$ that is proportional to
the mean excess radial component $E_\mathrm{ex} = E_\rho - E_\mathrm{f}$
affecting the muon dynamics, namely
\begin{equation}
	\Omega^{\scriptscriptstyle \Delta E}_\theta = \frac{e}{m}
	\left( a - \frac{1}{\gamma^2 - 1}
	\right)\frac{\beta_z(t)}{c}E_\mathrm{ex}.
	\label{eq:precession_mdm_excess_efield}
\end{equation}
In a realistic scenario where the orbit centre is displaced from the $E$-field
central axis, the radial component $E_\rho$ would be position and momentum dependent. This is
explored in the next section~\ref{sec:e-field_uniformity}.

Taking into account Eq.~\eqref{eq:b_theta}, we can approximate the angular velocity of the spin precession along the azimuthal $\theta$-axis to \begin{equation}
	\Omega^{\,\scriptscriptstyle\beta\cdot B}_\theta =
	\frac{e}{m} \left(\frac{a\gamma}{\gamma + 1}\right) \beta_\theta^2 B_\theta(t),
	\label{eq:precession_mdm_theta}
\end{equation}
due to the second-order with respect to the velocity term $(\vec{\beta} \cdot \vec B)\vec{\beta}$  in the Thomas-BMT equation, where a possible residual magnetic field $B^\Kick_\theta$ may be the magnetic kick or induced eddy currents, similar as in Eq.~\eqref{eq:b_rho_arbitrary}.

In summary, we combine Eqs.~\eqref{eq:beta_z}, \eqref{eq:precession_mdm_excess_efield}, and~\eqref{eq:precession_mdm_theta}, and include an arbitrary
azimuthal $B$-field, $B^\Kick_\theta(t)$, for the total azimuthal angular velocity,
\begin{multline}
	\Omega^\MDM_\theta =
	\frac{e}{m}\frac{p_{z_0}}{p_\theta}
	\sin( \omega_\mathrm{b} t)
	\left[ \left( a - \frac{1}{\gamma^2 - 1} \right)
		\frac{\beta_\theta}{c}E_\mathrm{ex} - \right.\\ - \left(\frac{a\gamma}{\gamma + 1}\right) \beta_\theta^2
		B_z\bigg] -
	\frac{ea}{m} B^\Kick_\theta(t),
	\label{eq:precession_mdm_z_oscillations}
\end{multline}
due to the AMM\@.
Averaged over $t \gg \omega_\mathrm{b}^{-1}$ results in
\begin{equation}
	\left\langle \Omega^\MDM_\theta\right\rangle =
	-\frac{e a}{m} \langle B^\Kick_\theta(t) \rangle.
	\label{eq:precession_mdm_theta_avg}
\end{equation}

On the closed orbit, the average azimuthal magnetic field $\langle
	B^\Kick_\theta \rangle$ is equal to zero when no current flows through the area enclosed by
the muon orbit. Despite the fact that $\left\langle \Omega^\MDM_\theta\right\rangle=0$, it remains
informative to identify and quantify the primary sources of oscillations. Furthermore, when these oscillations are combined with those around the other two axes, it is possible that a geometric phase may accumulate, as discussed in section~\ref{sec:berry_phases}.

\subsection{Description of the electric field in the storage-ring region}\label{sec:e-field_uniformity}

The radial electric field $E_\rho$, essential for the frozen-spin condition, may not be constant along the muon orbit. This might occur if the axes of the electrodes are not aligned with that of the solenoid field  or if the muon orbit is not centred to the axes of the electrodes.

To determine the components of the electric field in the muon reference frame, we consider a purely radial electric field generated by perfectly coaxial cylindrical electrodes with radii $A$ and $B$ with $A<B$.
In the reference frame $(x', y', z')$ of the electrodes (see Fig.~\ref{fig:coord_system_a}),
\begin{equation}
	\vec E'(x', y', z') = \frac{V}{\log{\frac{B}{A}}}\begin{pmatrix}\frac{x'}{(x')^2 + (y')^2} \\ \\ \frac{y'}{(x')^2 + (y')^2}
		\\ \\ 0\\\end{pmatrix},
	\label{eq:e_field_vector}
\end{equation}
where $z'$ is parallel to the solenoid axis,
which  might be displaced with respect to the centre of the muon orbit.
By applying a rotation, $R(\delta)$, around the $y'$-axis by an angle~$\delta$,
we transform the electric field,
\begin{equation}
	\vec E = R_z(\delta) \vec E'(R_z^{-1}(\delta)\vec r + \vec r_0),
	\label{eq:e_field_ref_frame}
\end{equation}
in the rest frame of the muon, where $\vec r_0 = (x'_0, y'_0, 0)$ is the displacement between the centre of the muon orbit and the position of the the electrode's axis.
Due to the cylindrical symmetry of the $E$-field around its central axis, we can always select the reference frame such that arbitrary displacements can be represented in this manner.
Therefore, the electric field in the reference frame defined by the longitudinal magnetic field may be written as
\begin{equation}
	\vec E(x, y, z) = \frac{V}{\log{\frac{B}{A}}}
	\begin{pmatrix}
		\frac{x}{r^2} \cos\delta \\ \\
		\frac{y}{r^2}            \\ \\
		-\frac{x}{r^2} \sin \delta
	\end{pmatrix},
	\label{eq:e_field_rotated}
\end{equation}
where $y = y' + y'_0$, $x = x'_0 + x' \cos\delta - y' \sin\delta$, and $r^2 =
	x^2 + y^2$.

The average of the radial electric field over the circular orbit of the muon,
\begin{equation}
	\tilde E_\rho = \langle E_\rho \rangle_\phi =
	\frac{1}{2 \pi} \int_0^{2\pi} E_\rho(\rho, \phi, z) d\phi,
	\label{eq:average_e_field}
\end{equation}
can be obtained most easily by representing $\vec E(\vec r)$ in cylindrical coordinates and integrating over the angle $\phi$, shown in Fig.~\ref{fig:coord_system_a}.
To consider the spin motion due to the cyclotron motion
in the electric field,
we approximate the radial component,
\begin{equation}
	E_\rho(t) \approx \tilde E_\rho +
	\frac{1}{2}\left(E_{\rho,\rm max} - E_{\rho,\rm min}\right)\cos (\omega_\mathrm{c} t +
	b_0),
	\label{eq:model_e_field_cyclotron}
\end{equation}
where $E_{\rho, \rm max}$ and
$E_{\rho,\rm min}$ are the maximal and minimal values of the electric field along a
muon orbit and $b_0$ is the initial phase of the muon position along the orbit.

Note that Eq.~\eqref{eq:average_e_field} is valid only in the case of a
circular orbit. In this case, it can be shown numerically that
\begin{equation}
	\left\langle E(\rho, z) \right\rangle_\phi = \left\langle E'(\rho, z)\right\rangle_\phi,
	\label{eq:rotation_orbit_doesnt_matter}
\end{equation}
which means that a tilt of the concentric assembly of inner and outer electrodes with respect to the main $B$-field axis does not influence the average frozen-spin condition and, more
importantly, does not change the net $E_z$ component.
Another significant consequence of this finding is that displaced muon orbits would still experience the same average radial component as the
nominal orbit, ensuring that displacements do not influence the storage of muons.

As the centripetal force due to the $B$-field is about a factor $10^3$ larger than that due to the $E$-field, and the expected misalignment between the centre of the orbit and the centre of the inner electrode is small, the circular orbit approximation holds well.
Another source for non-circular orbits is a non-uniform magnetic field, $B_z$, which is discussed in section~\ref{sec:uniformities}.

\subsection{EDM-like spin precession}
The signature of an EDM is the time-dependent asymmetry between decay positrons emitted along or opposite the $B$-field ($z$-axis), which is proportional to the change in the projection of the spin along the $z$-axis.
The angular velocity of Eqs.~\eqref{eq:precession_mdm_total_rho} and~\eqref{eq:precession_mdm_z_oscillations} projected along the $z$-axis is
\begin{equation}
	\vec\Omega^\MDM\cdot \hat{z} = \Omega^\MDM_\rho\cos(\omega_z t + \Phi_0) +  \Omega^\MDM_\theta \sin(\omega_z t + \Phi_0),
	\label{eq:projection_omega_mdm}
\end{equation}
where
\begin{equation}
	\omega_z = -\frac{e}{m}\left[aB_z - \left( a + \frac{1}{1 - \gamma^2}
		\right)\frac{\beta}{c}\tilde E_\rho \right]
	\label{eq:omega_g-2}
\end{equation}
is the angular velocity of the precession around $z$ due to an imperfect cancellation of the $g-2$ precession and $\hat{z}$ is the unit vector along $z$. In reality, $E_\rho$
will oscillate, according to Eq.~\eqref{eq:model_e_field_cyclotron}, with
frequency $\omega_\mathrm{c}$, due to the changing distance between the muon and
the centre of the electrodes generating the electric field. $B_z$ will oscillate
with frequency $\omega_\mathrm{b}$ due to the variation of the weakly-focusing
field. In a well-tuned frozen-spin experiment, $\omega_z$ is much smaller than
$\omega_\mathrm{b}$ and $\omega_\mathrm{c}$, and the rotation of the spin around
$z$ can be approximated with a constant angular velocity using the average values
of $E_\rho$ and $B_z$. The total longitudinal rotation of the spin with
respect to the momentum is
\begin{equation}
	\Psi(t) = \int_0^t \vec\Omega_\MDM\left(t'\right)\cdot \hat{z}\,\diff t'.
	\label{eq:precession_theta_integral}
\end{equation}

To verify the validity of the equation derived for the total longitudinal
rotation corresponding to the EDM signal, we have set up a \texttt{Geant4} Monte
Carlo simulation of the storage
ring region of the experimental setup. While \texttt{Geant4} is not a usual
choice for storage ring simulations it was deemed optimal for this experiment as
the detectors, field generating elements (coils, electrodes), and muon orbit are
in close proximity and very much interlinked. Though the available integration
algorithms in \texttt{Geant4} are non-symplectic, the effects on the tracked
position, momentum, and spin direction are smaller than the effects from running
the simulations with \SI{1e-7}{mbar} air pressure within the experimental
volume.
Additionally, testing the code with asymptotically small step sizes in the range of \SIrange{0.01}{2.0}{mm} converged to a stable solution for small step size on the nominal orbit, indicating that the direction of the muon spin as a function of time did not show significant variations with changes in step size within this range.

The EM field in the simulation is read from field\-maps generated by finite element
simulations using the software \texttt{ANSYS} \texttt{Maxwell3D}~\cite{Ansys_ED}.
The effects of finite spacing between points on a regular grid on which the EM field is defined
have a significantly larger impact than the choice of integration scheme and
step size.
The optimisation of the EM field generation as well as the
verification of the simulation and derivation of analytical equations are
shown in detail in \ref{sec:g4_verification}.

\subsection{Spin precession due to geometric phases}\label{sec:berry_phases}

The geometric phase, also known as Berry's phase, is a phase difference acquired
over the course of a cycle in parameter space when the system evolves
adiabatically~\cite{Berry1984}.\footnote{Note that an equivalent effect exists also in classical mechanics~\cite{Hannay:1985}.}
Such cycles in the parameter space can occur due to the periodic oscillations of stored muons in the non-uniform electric and magnetic fields of the experimental device.
In classical parallel transport, the phase accumulation is equal to the solid angle subtended by a vector on the spherical surface in parameter space.
For quantum parallel transport in fermions, where the vector is the spin moving through the $B$-field space, the geometric phase is half of that~\cite{Sakurai2011}.

Let us assume that there are two oscillations around the perpendicular axes
$x$ and $y$ with a time dependent angular velocity in the form
\begin{equation}
	(\Omega_x, \Omega_y) = \left(A_x \cos(\omega_x t), A_y \cos(\omega_y t + b_0)\right).
	\label{eq:parametric_oscillations_0}
\end{equation}
Integrating the expressions with respect to time, the accumulated phase as a function of time is
\begin{equation}
	a_x(t) = \frac{1}{\omega_x}A_x \sin(\omega_x t),\,\text{and}~a_y(t) =
	\frac{1}{\omega_y}A_y \sin(\omega_y t + b_0),
	\label{eq:parametric_oscillations}
\end{equation}
where $\omega_x$ and $\omega_y$ are the angular frequencies of the oscillations,
$A_x$ and $A_y$ are the peak angular velocities of the
spin precession around the respective axis, and $b_0$ is the
difference in their phases at time $t = 0$, which corresponds to the end of the magnetic pulse used to store the muons on a stable orbit.
The peak angular velocities,
\begin{align}
	A_B & = -\frac{ea}{m}B_\mathrm{max},
	\label{eq:omega_b_max}                                       \\
	A_E & = \frac{e}{mc} \left(\frac{1}{\gamma^2 - 1} - a\right)
	\left(\vec\beta\times\vec E\right)_\mathrm{max},
	\label{eq:omega_e_max}
\end{align}
of the spin precession are
proportional to the amplitude of oscillation of the EM field in the reference frame of the particle.

In the case of small oscillations, the surface of the unit sphere can be approximated with a plane and the enclosed solid angle can be approximated with the area enclosed by the curves.
The area under parametric curves,
\begin{equation}
	\mathcal{A}(t) = \frac{1}{2}\int (a_x \dot a_y - a_y \dot a_x)dt,
	\label{eq:greens_theorem}
\end{equation}
is calculated using Green's theorem.
In the case where $\omega_x \neq \omega_y$ one obtains\hfill
\begin{strip}
	\rule[-1ex]{\columnwidth}{1pt}\rule[-1ex]{1pt}{1.5ex}
	\begin{multline}
		\mathcal{A}(t; \omega_x, \omega_y, b_0) = \frac{1}{2} \frac{A_x^\peak
			A_y^\peak}{\omega_x \omega_y} \int \left( \omega_y \cos(\omega_y t + b_0)\sin(\omega_x t)
		- \omega_x \cos(\omega_x t)\sin(\omega_y t + b_0)\right)dt = \\
		=\frac{1}{4}\frac{A_x^\peak
			A_y^\peak}{\omega_x \omega_y}\left[\frac{\omega_x - \omega_y}{\omega_x + \omega_y}\cos( (\omega_x + \omega_y)t + b_0) -
			\frac{\omega_x + \omega_y}{\omega_x - \omega_y} \cos( (\omega_y -
			\omega_x)t + b_0)\right],
		\label{eq:berry_phase}
	\end{multline}
	\hfill\rule[1ex]{1pt}{1.5ex}\rule[2.3ex]{\columnwidth}{1pt}
\end{strip}%
for the integral, which for resonant oscillations, $\omega =
	\omega_x = \omega_y$, is
\begin{equation}
	\mathcal{A}(t; \omega, b_0) = -\frac{t}{2\omega}A^\peak_x
	A_y^\peak \sin(b_0),
	\label{eq:berry_phase_equal_phases}
\end{equation}
resulting in an angular velocity
\begin{equation}
	\mathcal{\dot A}(\omega, b_0) = -\frac{1}{2\omega}A^\peak_x
	A_y^\peak \sin(b_0).
	\label{eq:berry_phase_equal_phases_derivative}
\end{equation}

Another approach to obtain Eq.~\eqref{eq:berry_phase_equal_phases} is by using the method of averages and performing a second-order approximation of the exact Thomas-BMT equation, as done in the works of Carli and Haj Tahar~\cite{HajTahar2021,Carli2021,Carli2022}.

By using equations~\eqref{eq:berry_phase} and~\eqref{eq:berry_phase_equal_phases} one can calculate the phase accumulation as a function of time in the case of two periodic oscillations along the
perpendicular axes.
It can be seen that the geometric phase becomes larger with
decreasing differences between the frequencies of the two oscillations.
In the case of equal frequencies, the phase accumulation is linear with time and proportional to the product of the peak angular velocities of the spin
precession around the two axes.
In the case of equal frequencies, the geometric phase is zero when the two
oscillations are in phase $(b_0 = 0)$ and is maximal when they are out of phase ($b_0 = \pi/2$).
The validity of geometrical phase calculations was verified by
Monte Carlo simulations and is presented in~\ref{sec:geometric_verification}.

An example of a potential geometric-phase effect for the muon EDM experiment is
resonant oscillations of the spin around the longitudinal and radial axes due to the cyclotron motion of muons in the electric field for the frozen-spin technique.
This can happen when the centre of the muon's orbit is offset from
the centre of the electric field, combined with an angular misalignment of the
axis of the coaxial electrodes and that of the solenoid field, as outlined in
Eq.~\eqref{eq:e_field_ref_frame}. The angular misalignment would lead to oscillations in $E_z$ in the rest frame of the muon and the offset of the orbit to a changing $E_\rho$.
Despite the null net $(g-2)$ precession over a cycle and null net precession due to the $E_z$ component in the muon reference frame, indicated in
Eq.~\eqref{eq:rotation_orbit_doesnt_matter}, small oscillations
around the $z$- and $\rho$- axes will occur at the cyclotron angular frequency
$\omega_\mathrm{c}$, potentially leading to a systematic effect discussed in~\ref{sec:geometric_verification}.

\subsection{Other sources of spin precession}
Other effects that could lead to a precession of the spin come from the muon
motion in the curvature of space-time due to Earth's gravitational field, and
a possible influence of synchrotron radiation. Although these effects are
negligible we have included their estimates for completeness.

\subsubsection{Gravity}
There are two effects of gravity that lead to a spin precession that could mimic an EDM\@. The direct contribution~\cite{Lszl2018,Pretz2020},
\begin{equation}
	\Omega_\mathrm{GR} = \frac{2\gamma
		+1}{\gamma+1}\frac{\beta}{c}g_\mathrm{e},
	\label{eq:omega_gr}
\end{equation}
results from general relativity, where $g_\mathrm{e}$ is the gravitational acceleration at the surface of the Earth.
The second contribution is due to an effective restoring force from
either the electric or magnetic field that prevents the particles from falling.
The $E$-field that is necessary to counteract the gravitational attraction of
the earth is
\begin{equation}
	E_\mathrm{g} = -\frac{2\gamma^2 - 1}{\gamma} \frac{m}{e} g_\mathrm{e}.
	\label{eq:efield_gr}
\end{equation}
The magnitude of $E_\mathrm{g}$ for both experimental phases is below
$\SI{30}{nV/m}$.

In the muon EDM experiment case, both the direct and indirect effects of gravity
lead to angular velocity of the spin precession on the order of \SI{10}{nrad/s},
or more than seven orders of magnitude below the statistical sensitivity. The
influence of gravity on the spin precession of muons in storage rings are also
calculated in~\cite{Kobach2016} and estimate the systematic effect at the same
order of magnitude. Therefore, we consider gravitational effects as negligible
and will not discuss them further.

\subsubsection{Synchrotron radiation}
The muons will lose energy when circulating in the storage ring due to
synchrotron radiation.
This will not lead directly to spin precession, but could result in depolarisation.
The power emitted by synchrotron radiation can be calculated by applying the relativistic Larmor formula,
\begin{equation}
	P_\gamma = \frac{1}{6\pi\varepsilon_0}\frac{e^4}{m^2c}\gamma^2\beta_\theta
	B_z.
	\label{eq:synchtrotron}
\end{equation}
For the Phase~I and Phase~II experiments this results in an average emission of \SI{1.46}{\micro eV/
	\micro s} and \SI{23.0}{\micro eV/ \micro s}, respectively.
Such rate of reduction in the muon kinetic energy is negligible and would not lead to any measurable effect.

Synchrotron radiation can also cause gradual longitudinal polarisation of the particles (Sokolov–Ternov effect) according to $P \approx 1 - e^{t/\tau_\mathrm{p}}$. The polarisation is perpendicular to both velocity and acceleration, thus along the magnetic field responsible for the bending. The characteristic time $\tau_\mathrm{p}$ is~\cite{Jackson1976}
\begin{equation}
	\tau_\mathrm{p} = \frac{8}{5\sqrt{3}} \frac{m^2c^2\rho_0^3}{e^2\hbar\gamma^5}.
\end{equation}
For the parameters of the Phase I and II experiments the characteristic time amounts to $\tau_\mathrm{p}\approx\SI{e20}{s}$, to be compared with the typical measurement time of \SI{e-5}{s}. Therefore, spin-flip synchrotron radiation is not a concern for the experiment.

\section{Spatial and temporal non-uniformity of the EM field}
In this section, we provide calculations for specific deviations from the ideal homogeneous EM-fields, EM-field non-uniformity, that lead to false EDM signals.
Such signals can be observed by AMM-induced spin precession around the radial or azimuthal axes.
The latter can occur if there is a non-zero azimuthal magnetic field component in the rest frame of the particle.
This requires a net current flowing through the area
enclosed by the muon orbit. All electric supply current leads are designed such that the net current flow is expected to be zero in the experimental setup.
Therefore, special attention is given to the sources of radial spin precession.
In this context the two most significant sources of systematic effects are: ({\it i}\,) a $z$-component of the electric field, and ({\it ii}\,) a time-varying radial $B$-field
component.

Another possible source of a false EDM signal that is explored is the effect of
a time-variable magnetic field that leads to a longitudinal shift of the average orbit position.
Finally, we derive limits on the deviations of the fields from
their nominal values and their orientations, specifying requirements for the
realisation of the experimental setup.

\subsection{Field non-uniformity}\label{sec:uniformities}
Muons on the nominal orbit experience only a $B$-field along the $z$-axis, of about
\SI{3}{T}, and a purely radial $E$-field, such that the effect of any anomalous magnetic moment is cancelled and the relative angle between spin and momentum is constant.
Any deviation of the fields from this ideal configuration or from the nominal orbit induces spin motion.

In the following analysis, we exploit that arbitrary motions can be represented as a sum of oscillations around mutually perpendicular axes to describe the effects of dynamic and geometric phases.
For oscillations with a period much shorter than the measurement time of several muon lifetimes, the mean of the dynamic phase around each axis would tend to zero; the accumulation of a geometric phase remains possible.

The phase accumulation due to geometric phases can be calculated using Eq.~\eqref{eq:berry_phase_equal_phases}. We can distinguish three types of geometric phases that can be observed: ({\it i}\,) due to oscillations in the spin direction caused by oscillations in the $B$-field along two perpendicular axes, ({\it ii}\,) due to oscillations in the spin direction caused by oscillations in the $E$-field along two perpendicular axes,  ({\it iii}\,) due to the coupling of oscillations in the $B$- and $E$-fields.
The phase accumulation due to oscillations with given frequencies can be
calculated using Eqs.~\eqref{eq:berry_phase} and
\eqref{eq:berry_phase_equal_phases} by substituting the $B$- or $E$-field
oscillation amplitude in Eq.~\eqref{eq:omega_b_max} or~\eqref{eq:omega_e_max}, respectively.
A concrete analysis in the case of the Phase~I muEDM experiment
is given in the discussion, section~\ref{sec:ebgeom_limits}.

For low-frequency oscillations $\omega \ll \tau_\mu^{-1}$ one has to
consider dynamic phase accumulation as well.
In this case, a systematic effect can occur if an oscillation of a field in the muon reference frame is correlated with the
injection time.
As the measurement variance scales with the number of detected
decay positrons, which will decrease exponentially with time, earlier times will
have larger weight on the final asymmetry.

For a measurement window $L$, the weighted average of an oscillation with unit
amplitude and angular frequency~$\omega$, weighted over the number of muons at a given time $t$ after injection is
\begin{equation}
	W(\omega) = \left(\int_0^L e^{-t/\tau_\mu^\ast}dt\right)^{-1} \int_0^L \cos(\omega t+b_0)e^{-t/\tau_\mu^\ast} dt,
	\label{eq:osc_weights}
\end{equation}
where the first multiplier on the left-hand side is a normalisation factor. The boosted muon lifetime
is $\tau_\mu^\ast = \gamma \tau_\mu$ with $\tau_\mu \simeq
	\SI{2.197}{\micro\second}$. For $L \gg \tau_\mu^\ast$ Eq.~\eqref{eq:osc_weights} reduces to
\begin{equation}
	W_\mathrm{L}(\omega; b_0) = \frac{\cos(b_0) + \gamma\tau_\mu\omega\sin(b_0)}{1+(\gamma\tau_\mu \omega)^2}.
	\label{eq:osc_weights_reduced}
\end{equation}
For all further analysis we assume $b_0 = 0$ and $W_\mathrm{L}(\omega) = W_\mathrm{L}(\omega; 0)$, as we are interested in low frequency signals that could mimic an EDM\@.

In the case of spin precession due to the AMM coupling with the radial component of a time-varying magnetic field field $B_\rho(z(t)) = \int_0^\infty A_\rho(\omega) \cos(\omega t + b_0) d\omega $ expressed as in Eq.~\eqref{eq:b_rho_arbitrary}, the requirement that the angular velocity is less than a fraction $F$ of the experimental sensitivity is
\begin{equation}
	\frac{ea}{m}A_\rho(\omega) W_\mathrm{L}(\omega)
	\leq F\dot\Psi, 
	\label{eq:b_field_uniformity_0}
\end{equation}
where $F \in (0, 1)$ is an arbitrarily chosen factor.\footnote{Here we use $F=1/4$ to allow for up to 16 independent systematic effects, each at a quarter of the statistical sensitivity. This ensures that the total systematic uncertainty remains on par with statistical uncertainty, given that uncertainties are combined quadratically in the final analysis. Such a choice is naturally arbitrary and depends on the requirements of the particular experiment.} The limit as a function of $d_\mu$ then becomes
\begin{equation}
	A_\rho(\omega) \leq
	F \frac{1}{W_\mathrm{L}(\omega)}\frac{2mc}{ea\hbar}\beta_\theta B_z d_\mu.
	\label{eq:b_field_uniformity}
\end{equation}
Calculations of the limit on the radial $B$-field in the rest frame of the  muon for the muEDM experiment are given in section~\ref{sec:ebgeom_limits}.

In cases where the radial B-field amplitude is too large muons will not be stored. In order to derive a limit on the maximum combination of amplitude and oscillation frequency, we approximate the muon motion due to the oscillating $B$-field by a harmonic oscillator, hence, the following relation between amplitude and position holds
\begin{equation}
	\frac{A_\rho(\omega)}{z_\mathrm{max}} = \partial_z A_\rho(\omega) =
	\frac{\omega^2}{\omega_\mathrm{c}^2}\frac{B_0}{\rho_0},
	\label{eq:not_stored_condition}
\end{equation}
similar as for the calculation of the betatron oscillations (see Eq.~\eqref{eq:omega_betatron_calc}), and assuming a constant gradient of $B_\rho$ along $z$.
Therefore, the amplitude of the oscillations of the radial $B$-field,
\begin{equation}
	A_\rho(\omega) \leq z_\mathrm{max} \frac{\omega^2}{\omega_\mathrm{c}^2}
	\frac{B_0}{\rho_0},
	\label{eq:condition_not_stored}
\end{equation}
in the muon rest frame corresponding to the maximum longitudinal displacement $z_\mathrm{max}$.
Thus, we obtain the conditions for which the muon will not be stored in
the storage ring and therefore will not contribute to the measurement signal.

\subsection{Longitudinal electric field and alternating injections}%
\label{sec:longitudinal_efield}

Analogously to the derivation of a limit on the phase accumulation due to dynamical phases in an oscillating radial $B$-field (Eq.~\eqref{eq:b_field_uniformity}), we can derive a limit,
\begin{equation}
	E_z(\omega) \leq F \left(\frac{e}{mc} \left(\frac{1}{\gamma^2 - 1} - a\right) W_\mathrm{L}(\omega) \right)^{-1}\frac{2c}{\hbar} B_z d_\mu
	\label{eq:e_field_uniformity}
\end{equation}
for that induced by a longitudinal $E$-field $E_z$ using Eq.~\eqref{eq:precession_ey}.

The most stringent limit on $E_z$ is reached at low frequencies  approaching a constant value for $\omega \rightarrow 0$.
This limit can be relaxed considerably by taking advantage of the CP-violating nature of the EDM.
Alternating periodically between clockwise (CW) and counter-clockwise (CCW) particle motion in the storage ring, with otherwise identical conditions, permits cancellation of systematic effects in the measured asymmetry arising from $E_z$-induced dynamical phase accumulation.
This can be achieved by switching the polarity of the currents generating the magnetic field, thus inverting the direction of the magnetic field, and correspondingly reversing the injection direction of the muons.

Being proportional to $\vec\beta\times\vec B$, the EDM signal maintains its sign and is unchanged between the alternating injection modes.
The systematic effect related to a $z$-component of the $E$-field is proportional to $\vec
	\beta \times \vec E$. As the $z$-axis is defined to be aligned with the direction of the main $B$-field, $E_z$ will change sign between injections and so too
the systematic effect. Thus, in the ideal setup, the systematic effect will be
cancelled by summing the CW and CCW signals, while the EDM signal will double. However, the spin-phase build-up due to the longitudinal electric
field might be different for the two injection modes for a variety of reasons
that we will explore in more detail here.

Expanding Eqs.~\eqref{eq:precession_mdm_total_rho} and
\eqref{eq:projection_omega_mdm}, the projection of the angular velocity due to a
net non-zero longitudinal electric field (along $z$) that is less than a fraction $F$ of
the experimental sensitivity is
\begin{multline}
	-\frac{ea}{mc} \left(1 - \frac{1}{a(\gamma^2 - 1)} -
	\frac{1}{\beta_\theta^2}\right) \beta_\theta \lvert  E_z \rvert
	\cos(\omega_z t + \Phi_0)  \biggr\rvert_\CCW^\CW
	\\ \leq F\frac{2c}{\hbar}\beta_\theta B_z d_\mu.
	\label{eq:longitudinal_e_systematic}
\end{multline}
The evaluation bar denotes that we take the difference of the CW and CCW
signal, where the parameters $\gamma, \beta_\theta,
	E_z, \omega_z$ and $\Phi_0$ take the values corresponding to the two injection
modes. Note that $\beta_\theta$ and $\gamma$ are functions of the muon momentum
$p$, and $\omega_z = \omega_z(B_z, E_\rho, p)$.

As a consequence, four parameters need to be kept under strict control between
CW and CCW injections to fully cancel the false signal: the particle momentum
distribution for CW and CCW; the spin precession angular velocity around $z$,
which is proportional to a linear combination of $B_z$ and $\beta_\theta
	E_\rho$; the average initial phase $\Phi_0$ of the spin in the transverse plane
over the ensemble of injected particles; the average longitudinal component
$E_z$ along the CW and CCW trajectories. The false EDM depends on the product of
these parameters, therefore one cannot constrain a given parameter independently
of the others. Specific constraints are discussed in
section~\ref{sec:discussion}.

To cancel the effects of $E_z$ by alternating the direction of circulation of the
muons, it is necessary to ensure that on average the particles experience
similar $E_z$ at a given time after injection.
This requirement not only constrains the time stability and spatial uniformity of the applied electric field, but also the initial position and time evolution of the muon orbit which define the trajectory occupied within the field.

A possible source of time dependent changes of the muon orbit comes from the weakly-focusing field.
The simplest configuration of such field comprises a single current
loop, where the current flows in the opposite direction to that of the main
solenoid. This arrangement generates a gradient $\partial_z B_\rho$, which is
used to store muons in the $z$-direction. In conjunction with this, a radial
gradient $\partial_\rho B_z$ arises, leading to a variation of the longitudinal
$B$-field as a function of the distance from the centre of the weakly-focusing
coil.

If the centre of a particle's orbit deviates from the coil centre, the particle
will encounter a stronger field and a smaller turning radius in one portion of
the orbit and inversely on the opposite side. Consequently, this generates a
minor phase accumulation around the centre of the focusing field with each
cyclotron revolution (illustrated in Fig.~\ref{fig:orbit_precession}). This is
an instance of a magnetron oscillation,
\begin{equation}
	\omega_\mathrm{m} = \frac{\omega_\mathrm{b}^2}{2\omega_\mathrm{c}},
	\label{eq:magnetron}
\end{equation}
which is well described for Penning traps~\cite{Brown1982}.

Interestingly, the magnetron oscillation can be thought of as being caused by
the difference between the cyclotron and horizontal betatron\footnote{In the
	context of Penning traps the horizontal betatron frequency is commonly
	referred to as the reduced cyclotron frequency.} oscillations
\begin{equation}
	\omega_\mathrm{c} - \omega_\mathrm{h} = \omega_\mathrm{c}(1 -
	\sqrt{1-n}) = \omega_\mathrm{c}\frac{n}{2} + \mathcal{O}(n^2) =
	\frac{\omega_\mathrm{b}^2}{2\omega_\mathrm{c}} + \mathcal{O}(n^2),
	\label{eq:magnetron_alt}
\end{equation}
where for $n \ll 1$ the higher order terms in the Taylor expansion can be neglected.

The magnetron oscillation does not directly generate a false EDM signal,
but it could lead to different positioning of CW and CCW orbits. If there
is a significant field component $E_z$, the averages over time seen by the muon for the two injections might be unequal, thus failing to cancel the systematic effect. This
is discussed in greater detail in section~\ref{sec:efield_limits}.


\subsection{Longitudinal shift of the average orbit}%
\label{sec:longitudinal_shift}
The systematic effects discussed so far concern the precession of the spin
within the muon's reference frame. However, radial $B$-fields, external to the
weakly-focusing field, could lead to a rotation in the momentum vector, thereby
generating an EDM-like signal. The magnitude of this effect can be significantly
higher than the spin precession by approximately $(\gamma a)^{-1}$ times
($\approx$\,800 for the Phase~I experiment).

In the storage ring, the particle orbit is in a weakly-focusing field
characterised by a $\partial_z B_\rho$ gradient. While a constant external radial
$B$-field would merely alter the $z$-equilibrium position of the orbit and not cause a systematic effect, a radial field with amplitude $B'_\mathrm{tr}$ fluctuating in time could result in a drift of position, generating a systematic effect. The maximum amplitude of this longitudinal drift $d_\mathrm{l} = B'_{\text{tr}} / \partial_z B_\rho$, where the prime denotes
that the $B'_\mathrm{tr}$ is the magnetic flux density in the laboratory
reference frame.

The motion of the muon,
\begin{equation}
	\ddot z = -\frac{e}{\gamma m} c \beta_\theta \left[ z \partial_z B_\rho +
		B'_\mathrm{tr} \cos(\omega t) \right],
	\label{eq:external_b}
\end{equation}
depends on the combined effect of the weakly focusing field $ \partial_z B_\rho$
and the transient field $B'_\mathrm{tr} \cos(\omega t)$
with the oscillation frequency $\omega=2\pi f$.
The solution of \eqref{eq:external_b},
\begin{equation}
	z(t) = \frac{u B'_\mathrm{tr}}{u
		\partial_z B_\rho + \omega^2} \cos(\omega t) + z_0 \cos(\omega_\mathrm{b}t + \phi_0),
	\label{eq:external_b_solved}
\end{equation}
is similar to Eq.~\eqref{eq:omega_betatron_calc}, where $u = \frac{e}{\gamma m} c \beta_\theta$.
Considering only the first term with oscillations due to the transient field, and substituting $\omega_\mathrm{b} = u\partial_z B_\rho$, it is convenient to define the longitudinal pitch angle,
\begin{equation}
	P_\mathrm{l} = \frac{\dot z_\mathrm{tr} (t)}{c\beta_\theta} = - \frac{e}{\gamma m}\frac{\omega
	}{\omega_\mathrm{b}^2 + \omega^2}B'_\mathrm{tr}\sin(\omega t),
	\label{eq:external_b_velocity}
\end{equation}
as the ratio of longitudinal to azimuthal velocity. The time derivative,
\begin{equation}
	\dot P_\mathrm{l} = -
	\frac{e}{\gamma m}\frac{\omega^2
	}{\omega_\mathrm{b}^2 + \omega^2}B'_\mathrm{tr}\cos(\omega t),
	\label{eq:external_b_velocity_rate}
\end{equation}
yields the rate of change relevant for calculating the potential systematic effect.
We calculate an upper limit for this effect due to longitudinal drift of the muon orbit, by taking the weighted average of $\dot P_\mathrm{l}$,
\begin{equation}
	\dot P_\mathrm{W} = \frac{e}{\gamma m}\frac{\omega^2
	}{\omega_\mathrm{b}^2 + \omega^2}B'_\mathrm{tr}W_\mathrm{L}(\omega) \leq
	F\frac{2c}{\hbar}\beta_\theta B_z d_\mu.
	\label{eq:limit_ext_b_0}
\end{equation}
This results in a limit,
\begin{equation}
	B'_\mathrm{tr} \leq F \left[ \frac{e}{\gamma m}\frac{\omega^2}{\omega_\mathrm{b}^2 + \omega^2} W_\mathrm{L}(\omega) \right]^{-1}
	\frac{2c}{\hbar}\beta_\theta B_z d_\mu,
	\label{eq:limit_ext_b}
\end{equation}
as a function of $\omega$ and the betatron frequency
$\omega_\mathrm{b}$.
Specific limits for this orbit-drift-effect, which could result from the magnetic kick and thereby induced eddy currents are derived and discussed in section~\ref{sec:discussion}.

Note that Eq.~\eqref{eq:limit_ext_b} gives the limit on $B'_\mathrm{tr}(\omega)$
for a specific $\omega$. The residual tail from the magnetic kick will contain
a wide frequency spectrum. Therefore, the spin rotation due to the integral over
$\omega$,
\begin{equation}
	\frac{e}{\gamma m}\int_0^\infty
	\frac{\omega^2
	}{\omega_\mathrm{b}^2 + \omega^2}B'_\mathrm{tr}(\omega)W_\mathrm{L}(\omega)\, d\omega \leq
	F\frac{2c}{\hbar}\beta_\theta B_z d_\mu,
	\label{eq:limit_ext_b_integral}
\end{equation}
has to be constrained.

We can also derive limits on $B'_\mathrm{tr}(\omega)$ such that
$$\max(z_\mathrm{tr}(t)) \leq z_\mathrm{max},$$ where $z_\mathrm{max}$ is
defined in section~\ref{sec:uniformities} as the maximum longitudinal displacement for stored muons, similar to Eq.~\eqref{eq:condition_not_stored} that gives the relationship
between $B$-field oscillation amplitude and frequency in the muon rest frame for
stored muons.
With Eq.~\eqref{eq:external_b_solved} for $\cos(\omega t) = 1$ and
$\cos(\omega_\mathrm{b} t + \phi_0) = 1$, we obtain
\begin{equation}
	B'_\mathrm{tr} \leq
	(z_\mathrm{max} - z_0) \frac{\gamma m }{ec\beta_\theta}
	(\omega_\mathrm{b}^2 + \omega^2).
	\label{eq:zmax_external_b}
\end{equation}
For low frequencies $\omega \rightarrow 0$ the equation reduces to $z_\mathrm{max} - z_0 \geq
	B'_\mathrm{tr} / \partial_zB_\rho$, leading to the expected result that the
control of the maximum displacement due to external radial fields can be
established through the strength of the weakly focusing field.

\section{Discussion}\label{sec:discussion}
We have identified and studied several severe systematic
effects that, if not properly controlled, would limit the maximum achievable sensitivity for the muon EDM experiment.
These are effects related to the accumulation of geometrical phases, a non-zero average electric field along the
solenoid axis in the muon reference frame,
a transverse precession of the orbit around the axis of symmetry of the weakly-focusing field,
and a movement of the
average orbit due to a changing external radial magnetic field.

All calculations were performed assuming the parameters of the Phase~I muon EDM
experiment, unless explicitly stated otherwise. Phase~I aims to achieve a sensitivity of $\sigma(d_\mu)= \SI{3e-21}{\ecm}$.
We choose to set limits for each individual systematic effects to a fraction of the experimental sensitivity, $F = 1/4$.
The muon momentum is $p =
	\SI{28}{MeV/\textit{c}}$ corresponding to $\beta_\theta = 0.26$.
The main magnetic field is $B_z = \SI{3}{T}$, and the weakly focusing field has a gradient $\partial_zB_r = \SI{80}{mT/mm}$ at the radius of the nominal orbit.
Throughout this discussion we assume a worst-case scenario where all time-dependent systematic effects are correlated to the injection time and the initial parameters (polarisation, momentum, etc.) have a systematic offset between CW and CCW injections.

\subsection{Limits on geometrical phases and EM field non-uniformity}%
\begin{figure}[htb]
	\includegraphics[width=\columnwidth]{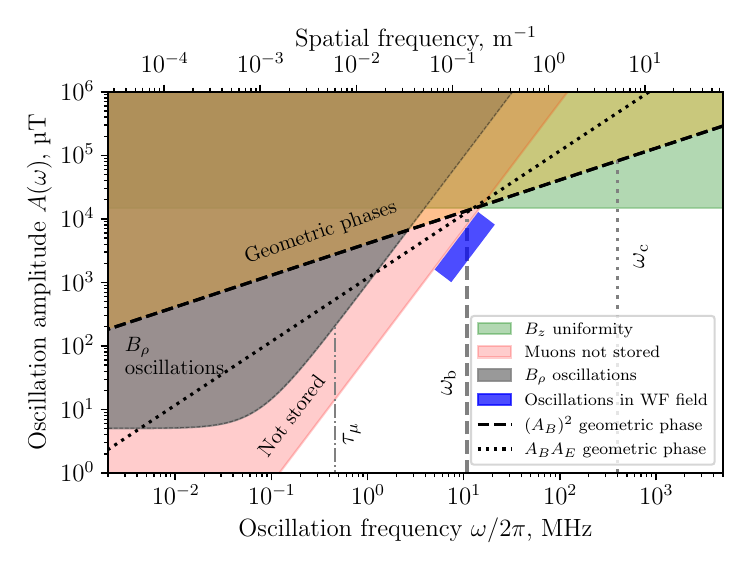}
	\caption{Limits deduced for a worst case false EDM signal in the Phase~I experiment
		due to resonant $B$-field oscillations as a function of the oscillation
		frequency and its amplitude (maximum deviation from the nominal) in the
		muon reference frame. The dashed vertical line shows the expected
		angular velocity corresponding to the betatron oscillations
		(\SI{150}{ns} period) and the dotted one to the cyclotron oscillations
		(\SI{2.5}{ns} period).
		The blue rectangle shows the possible values for the weakly-focusing field
		oscillation amplitude and frequency range.  The second abscissa shows
		the spatial frequency, giving the number of periods per metre travelled by
		the muons and calculated as $f = \omega / (2\pi \beta c)$.}
	\label{fig:berry_b_limit}
\end{figure}

\label{sec:ebgeom_limits}
We first deal with geometrical phases induced by non-uniformity in the
$B$-field according to the framework set up in section~\ref{sec:uniformities}.
To place limits on the field non-uniformity inducing geometrical phases we
assumed the worst case where the oscillations are maximally out of phase and the oscillations around the two perpendicular axes have the same amplitude.
The limits on the amplitude of the oscillation as a function of the oscillation frequency are shown in Fig.~\ref{fig:berry_b_limit} (depicted as the coloured
area above the $(A_B)^2$ dashed line).
The calculations were performed using
Eq.~\eqref{eq:berry_phase_equal_phases_derivative} requiring that the rate of
geometrical phase accumulation is a fraction $F$ of the experimental sensitivity
\begin{equation}
	\mathcal{\dot A}(\omega) = \frac{A_x A_y}{2\omega} \sin(b_0)  \leq
	F\frac{2c}{\hbar}\beta_\theta B_z d_\mu,
	\label{eq:limit_berry_bb}
\end{equation}
where $A_x = A_y = -eaB_\mathrm{max}/m$ and the initial phase $b_0 = \pi/2$.
For reference, the region of expected betatron oscillation amplitudes and frequencies is presented with a blue rectangle.

The geometric phase accumulation due to a combined $B$-field and $E$-field non-uniformity is shown in the
same figure with a dotted line labelled $A_B A_E$ at the level of 0.5\% of the radial
$E$-field required for the frozen spin state, corresponding to $E_\mathrm{max} = \SI{1.4}{kV/m}$.
The limit was calculated using Eq.~\eqref{eq:limit_berry_bb}, where $A_x =
	-eaB_\mathrm{max}/m$ and $A_y$ is given by Eq.~\eqref{eq:omega_e_max}.

The main source of a time-variable radial magnetic field is the split coil pair used to kick the particles into a stable orbit.
The nominal current pulse is a half-sine pulse with
\SI{100}{ns} half-period, producing a radial
magnetic field of a few hundred \unit{\micro\tesla} peak in the storage region.
A real pulse will not follow exactly the half-sine and will exhibit ringing with a finite decay time.
Another source of a slowly decaying radial magnetic-field components could be eddy currents induced
by this pulse in the electrode system or the bore of the main solenoid.

For low-frequency oscillations, especially when tending to zero, one has to
consider dynamic phase accumulation, where the limits on the amplitude of
the $B$-field oscillations are given by Eq.~\eqref{eq:b_field_uniformity}.
The limits from this source of systematic effects are given in
Fig.~\ref{fig:berry_b_limit} with a grey area labelled ``$B_\rho$
oscillations''.
The same reasoning can be applied to the azimuth
component of the magnetic field, but the expected amplitude of the $B_\theta$ oscillations is
negligible.

Not all combinations of oscillation frequency and amplitude of the
$B$-field lead to stored muons.
Using Eq.~\eqref{eq:condition_not_stored} and
imposing that the oscillation amplitude of the muon $z_\mathrm{max} \le
	\SI{50}{mm}$ we obtain the red exclusion area (labelled ``Not stored'') in
Fig.~\ref{fig:berry_b_limit}.
The \SI{50}{mm} limit on the longitudinal
oscillation amplitude is due to the positioning of the split coil pair of
the magnetic kicker.
Particles would only be able to be stopped within the
region where the radial magnetic field generated by the current pulse is such that the
Lorentz force $e c \vec\beta \cross \vec B$ counteracts the longitudinal motion,
which is in between the two current loops.

Systematic effects can only be caused
by $B_\rho$-field components, when limiting our considerations to the $B$-field, as only these will lead to an EDM-like spin precession. However, dynamic phase effects related to the electric field can be significant if the muon orbit deviates from circular and the
electrode system is tilted with respect to the central solenoid axis.
In this case, the
mean longitudinal component of the electric field over a cyclotron rotation would not be zero, as concluded from
Eq.~\eqref{eq:rotation_orbit_doesnt_matter}.
Assuming a \SI{1}{mrad} tilt of the
electrode system, the eccentricity, $e=\sqrt{1-a^2/b^2}$, where $a$ and $b$ and are the semi-major and semi-minor axis of the ellipse, of the muon orbit has to be kept below $e\leq0.1$.
This corresponds to a uniformity of the $z$-component of the $B$-field within
$\pm\SI{15}{mT}$ (depicted as ``$B_z$ uniformity'' in
Fig.~\ref{fig:berry_b_limit}). Effects of $B_z$ non-uniformity resulting in, e.g.,
magnetron oscillation are discussed in section~\ref{sec:longitudinal_efield}.

\begin{figure}[hbt]
	\includegraphics[width=\columnwidth]{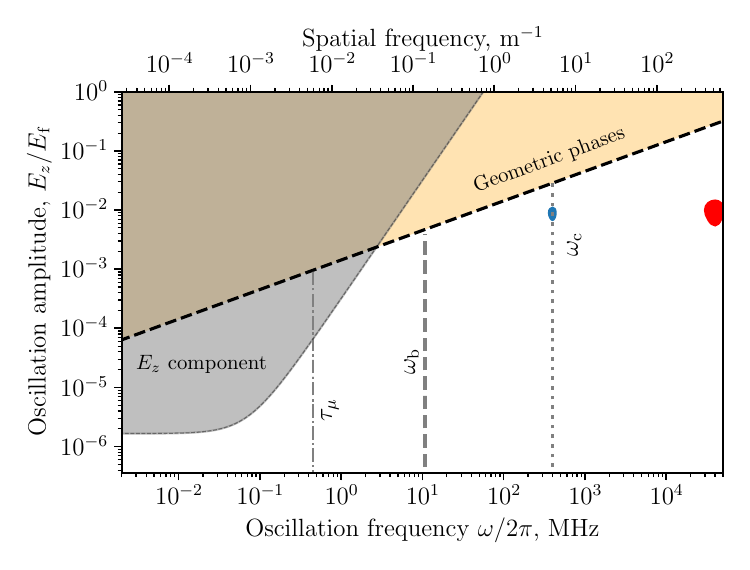}
	\caption{The worst case limits on the $E$-field oscillation frequency and amplitude (maximum
		deviation from the nominal) given as a fraction
		of $E_\mathrm{f}$. The dashed vertical line shows the angular velocity
		corresponding to the betatron oscillations and the dotted
		one to the cyclotron oscillations. The dashed-dotted vertical line corresponds to the
		muon lifetime and, coincidentally, the $g-2$ precession frequency without
		frozen-spin. The grey area is the limit of
		the longitudinal $E$-field as a result of the dynamical phase accumulation. The blue and red shapes show the
		parameter space for tilted electrodes and displaced
		muon orbit for solid or striped electrodes, respectively.}
	\label{fig:berry_e_limit}
\end{figure}

From the results shown in Fig.~\ref{fig:berry_b_limit}, we can see that the
geometric phases that are caused by $B$-field variation in time impose weaker
limits compared to other uniformity considerations. The observation of a
false EDM-signal due to oscillations of the spin around the radial axis is also
not possible, as the oscillations with significant amplitude would not
correspond to stored muons.
The unavoidable betatron oscillations due to the
weakly-focusing field do not evade the calculated constraints and will not lead to a significant false EDM-signal.

Nevertheless, the presence of a radial
$B$-field that is external to the weakly-focusing field would cause a shift in
the longitudinal position of the average muon orbit.
The systematic effects
related to a possible shift of the orbit equilibrium position with time are expanded upon in section~\ref{sec:external_mag_field}.

Figure ~\ref{fig:berry_e_limit} shows the limit derived from the geometric phase due to oscillations of the radial and longitudinal
electric-field components in the muon reference frame, using Eq.~\eqref{eq:limit_berry_bb}, where $A_x = A_y = A_E$ is defined by~\eqref{eq:omega_e_max}.

The expected false EDM signal due to $E_z$ oscillations was calculated using
Eq.~\eqref{eq:e_field_uniformity}.
The limit on $E_z$ uniformity is shown in Fig.~\ref{fig:berry_e_limit} as grey area.
All limits were calculated considering only a single
direction of circulation of the muons, i.e., only CW or CCW\@.


The analysis of the electric field uniformity shows that the maximum allowed
non-uniformity at the betatron frequency is $0.4\%$ of the electric
field $E_\mathrm{f}$ required to freeze the spin to the momentum.
Such a
time-dependent variation of the electric field in the muon reference frame can
occur due to the fringe field from the end regions
of the electrodes.
Studies using finite-element methods~(FEM) show that this effect can be mitigated by using sufficiently long, i.e.\ \SI{500}{mm}, electrodes, which would result in negligible fringe fields ($E_z \le \SI{0.02}{V/m}$) in the storage region.

The resonance between radial and longitudinal $E$-field oscillations caused by a
tilt in the electrode system and a displacement of the muon orbit could cause
the build up of geometric phases (discussed also in
section~\ref{sec:berry_phases}.
Assuming a \SI{1}{mrad} tilt of the electrodes
with respect to the central axis of the solenoid the muons will experience an
oscillating field $E_z$ at the cyclotron frequency with amplitude $\SI{0.3}{kV}$
due to the projection of the radial electric field along the $z$-axis.
If the orbit center is displaced from the central axis of the electrodes, the muons will also experience an oscillating radial $E$-field at the cyclotron frequency. The amplitude of the oscillation depends on the magnitude of the displacement and is approximately \SI{10}{kV} for \SI{1}{mm} displacement.
One can show that the combined effect of the tilt and displacement results in an equivalent geometric phase
build-up as if both oscillations would have an amplitude of \SI{1.7}{kV}.
This equivalent case (\SI{1.7}{kV} amplitude at $\omega_\mathrm{c}$) is contained in the blue
region in Fig.~\ref{fig:berry_e_limit} (lower limit of $E_z / E_\mathrm{f}$). Orbit displacement of \SI{3}{mm} would result in \SI{30}{kV} radial $E$-field oscillation, which, combined with the \SI{0.3}{kV} $E_z$ oscillation, is equivalent to both having \SI{3}{kV} amplitude. This case is also contained in the blue region (upper limit of $E_z / E_\mathrm{f}$). Note that both the tilt and the displacement used for the calculation are larger than the limits imposed by consideration of other sources of systematic effects.
This means that even in the worst-case scenario the accumulation of geometric phase due to a displacement of the muon orbit with respect to the centre of the electric field is negligible.

Another scenario for the generation of geometric phases is explored by assuming
the same tilt, but an electrode system composed of individual discrete wires instead of
a solid cylinder, discussed in \ref{sec:striped_electrodes}. This would
create non-uniformity in the radial electric field, which, in the muon reference
frame, will oscillate with a multiple of the cyclotron frequency. The red shape in Fig.~\ref{fig:berry_phases_efield} shows the additional parameter space occupied in this scenario, due to the additional source of geometric phase accumulation at the given multiple of $\omega_\mathrm{c}$. As the limit on field amplitude of the geometric phase effect increases with increasing frequency, a setup with a segmented electrode
is not excluded due to systematic considerations.
Potential benefits of this
electrode type are also explored in \ref{sec:striped_electrodes}.

\subsection{Systematic limits on a longitudinal electric field}%
\label{sec:efield_limits}
%

\begin{figure*}[htb]
	\centering
	\subfloat[]{\includegraphics[width=0.526\linewidth]{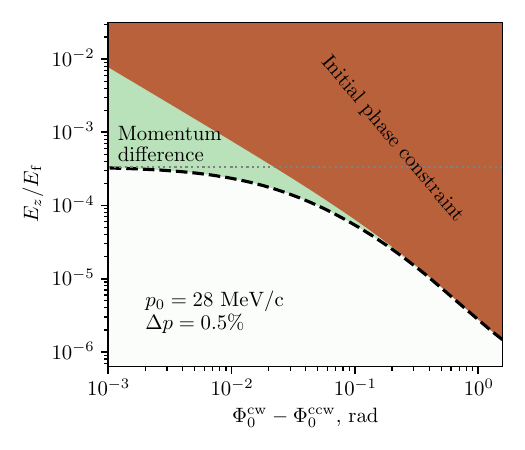}\label{fig:omz_ez_limit_a}}
	\subfloat[]{\includegraphics[width=0.451\linewidth]{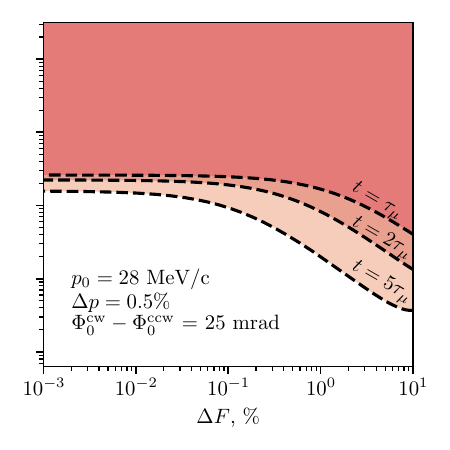}\label{fig:omz_ez_limit_b}}
	\caption{ Limit on the longitudinal $E$-field, $E_z$, when considering alternating
		CW/CCW injections, where the
		difference in mean momentum averaged over all injected muons for injections of CW and CCW beams is
		fixed at $\Delta p = 0.5\%$. a)~Shows the limit as a function of the initial phase difference
		$\Phi_0^\mathrm{cw} - \Phi_0^\mathrm{ccw}$. The horizontal dashed line indicates the limit coming from $\Delta$, keeping all other parameters equal between the two injection modes. The brown area is the
		constraint coming from the difference in initial phase at $\Delta p =
			0\%$. The green area and its thick dashed line edge show the combined
		limit of the two effects. b)~Shows the limit as a function of the difference in $E$- or $B$-field for CW
		and CCW orbits, calculated at various times.
		$\Delta F = 2 (F^\mathrm{cw} -
			F^\mathrm{ccw})/(F^\mathrm{cw} + F^\mathrm{ccw})$, where $F \in
			\{B_z, \tilde E_\rho \}$, while the difference in initial phases was fixed at
		\SI{25}{mrad}. }
	\label{fig:omz_ez_limit}
\end{figure*}

By far the main source of a systematic effect in the frozen-spin technique would
come from a non-zero $z$-component of the electric field. While this can be
significantly mitigated by employing counter-rotating beams, it follows from Eq.~\eqref{eq:longitudinal_e_systematic} that the combination of four parameters need to be kept constant between CW and CCW injections:
({\it i}\,) the particle velocity $\beta_\theta$, ({\it ii}\,) the spin precession angular velocity around
the longitudinal axis $\omega_z$, which is proportional to the linear
combination of $B_z$ and $\beta_\theta E_\rho$, ({\it iii}\,) the average initial phase
$\Phi_0$ of the spin in the transverse plane over the ensemble of injected
particles, and ({\it iv}\,) the average longitudinal component $E_z$ along the CW and CCW
trajectories. Note that the overall systematic effect is proportional to the
product of these parameters. Therefore, improvements in the control of a single
parameter are effective up to a point, after which one must constrain the rest
as well. For example, even controlling $\beta_\theta^\mathrm{cw}$ and
$\beta_\theta^\mathrm{ccw}$ to an excellent precision, the effect of a non-zero $E_z$
could still be significant due to differences between $\Phi_0^\mathrm{cw}$ and
$\Phi_0^\mathrm{ccw}$.

To untangle the problem, first we assume that $\langle E_z^\mathrm{cw} \rangle =
	-\langle E_z^\mathrm{ccw}\rangle = E_z$. Then we express $E_z$ from
\eqref{eq:longitudinal_e_systematic} as:
\begin{equation}
	E_z = \frac{F \dot\Psi}{-\frac{ea}{mc}\left[\left(
			1 - \frac{1}{a(\gamma^2 - 1)} - \frac{1}{\beta^2_\theta}
			\right)\beta_\theta \cos(\omega_z t + \Phi_0)\right]
		\Big|_\mathrm{ccw}^\mathrm{cw}},
	\label{eq:ez_cw_ccw_defined}
\end{equation}
giving the maximum permissible $E_z$ that would lead to a false EDM signal equal to the threshold of $F = 1/4$ of the experimental sensitivity.

The first parameter $\beta_\theta$ places constraints on the level of control of
the difference in momentum $\Delta p = p^\CW - p^\CCW$ for the two injection
schemes. For the Phase~I muon EDM experiment, using \SI{28}{MeV/\textit{c}}
surface muons, it is reasonable to aim for momentum control that ensures no more
than $\Delta p = 0.5$\% difference in the mean value of the momentum for injections of CW
and CCW beams. Note that such a difference not only leads to a difference in $\beta_\theta$, but also in $\omega_z$. Thus, we need to specify the
limit on $E_z$ at a given time $t$. Here we conservatively choose a time around
the end of the measurement $t =
	5\tau_\mu$ or \SI{11}{\micro\s}. With this constraint, it is possible to place a limit on the
difference between the initial phases $\Phi_0^\mathrm{cw}$ and
$\Phi_0^\mathrm{ccw}$. The maximum
permitted longitudinal $E$-field as a function of the difference of initial
phases is shown in Fig.~\ref{fig:omz_ez_limit_a}.
A further reduction of the initial phase difference below \SI{25}{\milli\radian} will become ineffective as
it reaches the limit set by the control of the momentum.

A determination of the initial phase can be achieved by a dedicated $g-2$ precession measurement, by observing the spin precession without
an electric field. We may tune the value of the initial phase using a Wien filter in the secondary beamline.
The stability of the radial electric field in the muon reference frame can be measured by applying $E=-E_\mathrm{f}$ to the electrodes and measuring the frequency stability of $\Omega^\MDM \simeq 2ea\lvert \vec B \rvert / m$.

The limits on $\Phi_0^{\mathrm{cw}} - \Phi_0^{\mathrm{ccw}}$ and $\Delta p$
constrain the initial condition for muon storage.
However, in the case where the spin is not perfectly frozen, the
spin phase will evolve with time. For the phase accumulation to remain the same
for both injections, the radial electric field in the muon reference frame must
be the same within some limits.
The same holds for the longitudinal $B$-field as
$\omega_z$ depends on the linear combination between $B_z$ and $\tilde E_\rho$.

The maximum longitudinal $E$-field component permissible in the muon reference
frame as a function of the difference in $B_z$ or $\tilde E_\rho$ is
shown in Fig.~\ref{fig:omz_ez_limit_b}, where $\Delta F = 2 (F^\mathrm{cw} -
	F^\mathrm{ccw})/(F^\mathrm{cw} + F^\mathrm{ccw})$, and $F \in \{B_z, \tilde E_\rho \}$.
The calculation was
performed at the fixed limit of the momentum and the initial phase difference
for the two injection schemes, $0.5\%$ and \SI{25}{mrad}, respectively.
The imperfect
cancellation of the systematic effect due to a longitudinal electric field is proportional to $\cos(\omega_z t)$, hence a function of storage time.
Assuming that the longitudinal electric field is strictly proportional to the radial $E$-field and hence the longitudinal
$B$-field, an improved control of better than $\Delta F \leq 0.01\%$ will not further reduce the effect, as the limiting factor becomes the initial phase and
the momentum difference.

Finally, under the assumption that the longitudinal $E$-field is unchanged between injection modes, we can place a limit of $E_z \leq 10^{-4}E_\mathrm{f}$.
This corresponds to $\Delta p = 0.5\%$, $\Phi_0^{\mathrm{cw}} - \Phi_0^{\mathrm{ccw}}
	= \SI{25}{mrad}$ and $\Delta \tilde E_\rho = 0.1\%$.
The stability of the $B$-field
can be controlled to an order of magnitude better than $\Delta B_z = 0.01\%$
(\SI{300}{\micro\tesla}) and does not contribute significantly to this limit. One
could tighten the limit on some parameter, in an attempt to relax the limit on
$E_z$, however, the contribution of the others will then become more
significant.
For example, reducing $\Delta p$ by an order of magnitude to 0.05\%
and $\Phi_0^{\mathrm{cw}} - \Phi_0^{\mathrm{ccw}}$ by 2.5~times to \SI{10}{mrad}
leads to a relaxed limit of $E_z \leq 5\times 10^{-4}E_\mathrm{f}$.

\subsubsection{Variation in the longitudinal electric field}
\begin{figure}[htb]
	\includegraphics[width=\columnwidth]{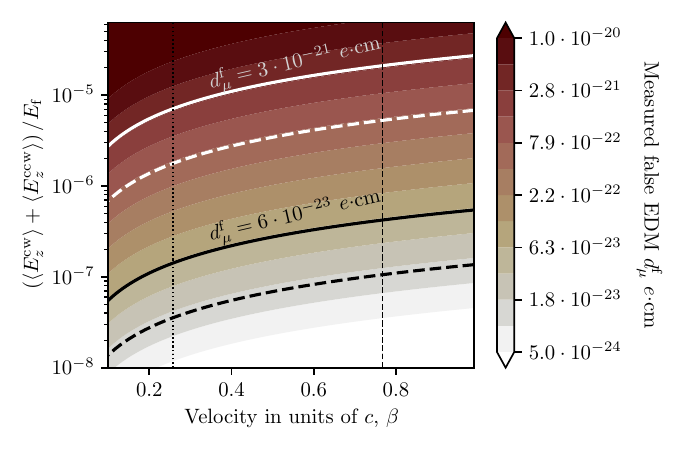}
	\caption{The measured false EDM $d_\mu^\mathrm{f}$ due to difference in the average
		electric field in the longitudinal direction between CW and CCW
		injections. The solid horizontal line shows the
		target sensitivity and the dashed horizontal line is
		one quarter of that value. The vertical dotted line is at $\beta =
			0.256$ (Phase~I) and the vertical dashed line is at $\beta =
			0.770$ (Phase~II).}
	\label{fig:ez_beta_limit}
\end{figure}

So far, we have assumed that the absolute value of the average $E_z$ over time and over all measured muons
is the same for the two injection modes, or $\langle E^\mathrm{cw}_z \rangle =
	-\langle E^\mathrm{ccw}_z \rangle$, which is only true if the muon
trajectories for CW and CCW overlap perfectly.
The false EDM signal measured as a function of the muon velocity and the
difference in the average longitudinal electric field in the muon reference
frame is shown in Fig.~\ref{fig:ez_beta_limit}.
For Phase~I of the
experiment this difference should be limited to $\langle E^\mathrm{cw}_z \rangle
	+ \langle E^\mathrm{ccw}_z \rangle \leq 2\times10^{-6}E_\mathrm{f}$.

Taking into account all considerations, we can place a limit on the maximum
$z$-component of the electric field allowed in the muon reference frame
of $E_z\leq 10^{-4}E_\mathrm{f}$, or approximately $\langle E_z^\mathrm{cw, ccw}
	\rangle  \leq$~\SI{28}{V/m} and its maximum change between the two injections $\left|\langle E^\mathrm{cw}_z \rangle + \langle
	E^\mathrm{ccw}_z \rangle\right|
	\leq \SI{0.56}{V/m}$ for the
Phase~I muon EDM experiment. The corresponding values for the Phase~II experiment are $\langle E_z^\mathrm{cw, ccw}
	\rangle  \leq$~\SI{2.9}{V/m} and $\left|\langle E^\mathrm{cw}_z \rangle + \langle
	E^\mathrm{ccw}_z \rangle\right|
	\leq \SI{0.15}{V/m}$.

\subsubsection{Slow drift of the muon orbit}~\\
One possible reason for differences between the absolute values of $\langle E_z^\mathrm{cw}\rangle$ and $\langle
	E_z^\mathrm{ccw} \rangle$ is the magnetron oscillation of the muon,
discussed in section~\ref{sec:longitudinal_efield}. For a typical Phase~I
cyclotron frequency of $\omega_\mathrm{c} = \SI{2.47}{rad/ns}$ and betatron
frequency $\omega_\mathrm{b} = \SI{0.073}{rad/ns}$, the magnetron frequency is
$\omega_\mathrm{m} = \SI{0.11e-3}{rad/ns}$ corresponding to a period of
\SI{5.8}{\micro\second}.

\begin{figure*}[hbt]
	\includegraphics[width=0.49\linewidth]{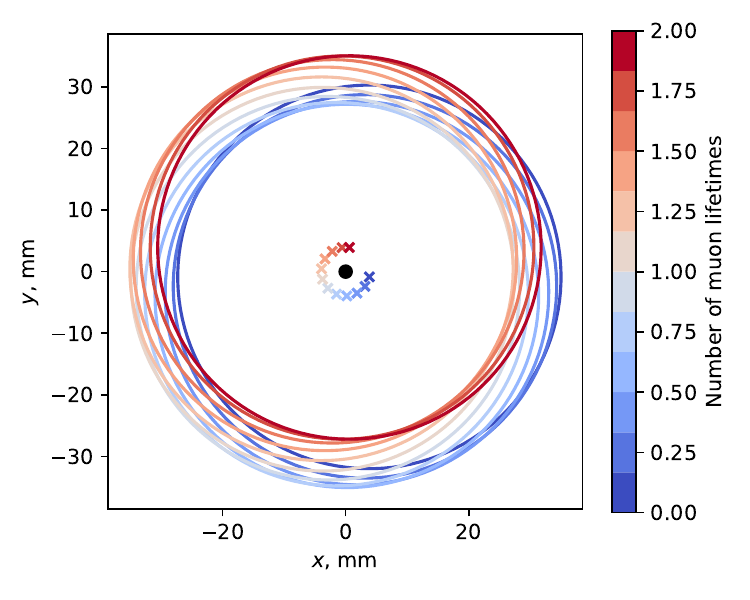}
	\includegraphics[width=0.49\linewidth]{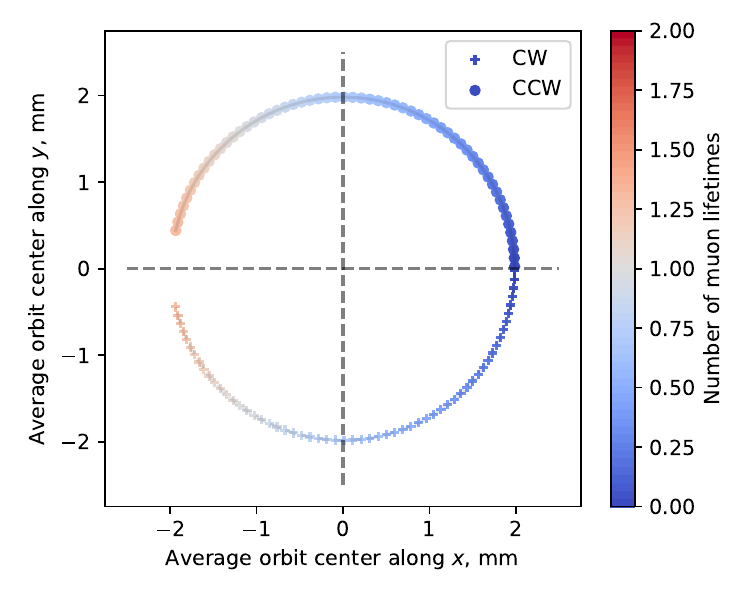}
	\caption{Precession of the muon orbit around the centre of the weakly-focusing field (magnetron oscillations).
		Left: Time evolution of the orbit of a single muon
		(line) and its centre (cross) around the centre of the focusing field
		(dot). Right: Precession of the mean centre of the orbits of an ensemble
		of muons for CW and CCW injections.}
	\label{fig:orbit_precession}
\end{figure*}

We explore this effect through a Monte Carlo simulation using \texttt{Geant4}.
The initial $z$-position of each particle was drawn from a Gaussian distribution with mean \SI{0}{mm} and \SI{10}{mm} standard deviation.
The transverse position was uniformly distributed along a circle with radius
equal to the nominal radius of the muon storage orbit, i.e.\ \SI{31}{mm}, and centred at $(x_0, y_0) = (2,0)\pm(3,3)$\,mm.
The momentum of the particles is sampled from Gaussian distribution with mean and standard deviation
$\SI{28\pm0.3}{MeV/\textit{c}}$. The sampling approximates conditions in which the muons are injected in the experiment slightly off the
nominal orbit. The simulations were performed with positive and
negative $B$-field to study the orbit precession for the CW and CCW
cyclotron motion. The mean position of the centre of the orbits as a
function of time is shown in Fig.~\ref{fig:orbit_precession}.

The observed magnetron oscillation period is consistent with the prediction of
\SI{5.8}{\micro\second}. From Eq.~\eqref{eq:magnetron}, the sign of the magnetron
oscillation follows the sign of $\omega_\mathrm{c}$, which can be seen in
Fig.~\ref{fig:orbit_precession} as well. The direction of the drift of the orbit
center changes between CW and CCW circulation. Another observation from the simulations performed is that, even though the initial muons start with random
displacements from the central axis of the weakly focusing field and a
distribution of momenta, the mean orbit centre follows a circular path with \SI{2}{mm} radius starting at the (2, 0)\,mm
position. Thus showing, that the mean position of the muon orbits at the beginning of the measurement (after the end of the nominal magnetic kick) is sufficient to
describe the mean magnetron motion.

The magnetron oscillation does not directly lead to a systematic effect, however, it might invalidate the assumption that $\langle E_z^\mathrm{cw}
	\rangle = - \langle E_z^\mathrm{ccw}\rangle$ as the muons in the two injection
schemes could sample a different volume and therefore a different longitudinal $E$-field.
This effect can be mitigated by ensuring that the mean centre of the
orbits coincides with the central axis of the weakly focusing field.
Another mitigation strategy is to limit the $\partial_x E_z$ and $\partial_y
	E_z$ gradients, such that the longitudinal electric field is sufficiently
uniform.
One can estimate the limits on those gradients by approximating
\begin{multline}
	\langle E^\mathrm{cw}_z \rangle = E^\mathrm{cw}_0 + \\ + \int_{0}^{2\pi}\left[(x_0 + \rho_0\cos \phi) \partial_x E_z + (y_0 + \rho_0 \sin \phi)\partial_y E_z \right]d\phi = \\ = E^\mathrm{cw}_0 + x_0\partial_x E_z  + y_0\partial_y E_z ,
\end{multline}
where $(x_0, y_0)$ is the offset of the orbit center from the central electrode axis and $E^\mathrm{cw}_0$ is a constant term in $E_z(\vec r)$. Then
\begin{equation}
	\langle E_z^\mathrm{cw}	\rangle + \langle E_z^\mathrm{ccw}\rangle = \delta_x\,\partial_x E_z + \delta_y\,\partial_y E_z ,
\end{equation}
where $\delta_x = \lvert x^\mathrm{cw}_0\rvert - \lvert x^\mathrm{ccw}_0\rvert$ and $\delta_y = \lvert y^\mathrm{cw}_0 \rvert - \lvert y^\mathrm{ccw}_0\rvert$, and noting that $E_0^\mathrm{cw} = -E_0^\mathrm{ccw}$ due to the coordinate axis $z$ following the $B$-field direction.

Assuming a maximum systematic offset between the mean centre of CW and CCW orbits of $\delta_x = \delta_y = \SI{1}{mm}$ implies that $(\partial_x E_z, \partial_y E_z) \leq \SI{0.56}{kV/m/m}$ for Phase~I. Given in terms of a fraction of the frozen-spin field this is $(\partial_x E_z, \partial_y E_z) \leq 0.2\% E_\mathrm{f}\unit{m^{-1}}$.

\subsection{Systematic limits on a transient radial magnetic field}\label{sec:external_mag_field}

As elaborated on in section~\ref{sec:longitudinal_shift}, a transient radial
magnetic field with angular frequency $\omega$ and amplitude $B'_\mathrm{tr}$
could introduce a systematic effect by rotating the momentum vector at a rate
$\dot P_\mathrm{l}$ around the radial axis, thus mimicking an EDM signal.
The constraints on a single frequency oscillating radial magnetic field with
oscillation amplitude $B'_\mathrm{tr}$ as a function of its angular frequency
$\omega$ are calculated using Eq.~\eqref{eq:limit_ext_b}, where we require that
$\dot P_\mathrm{W} \leq F \dot\Psi = \dot\Psi/4$, and where $\dot P_\mathrm{W}$ is the $\dot
	P_\mathrm{l}$ weighted over the exponential decay of the muons.

Using Eq.~\eqref{eq:zmax_external_b} we deduce a limit by requiring that the transient magnetic field does not lead to a too large  displacement of the muon along $z$ leading to a loss of the muons. The maximum amplitude of longitudinal oscillations, $z_\mathrm{max} =
	\SI{50}{mm}$, is defined by the maxima of the radial magnetic field in the weakly focusing field area.
A limitation of the maximum acceptable amplitude due to the betatron oscillations to
$z_0 = \SI{40}{mm}$ leaves $\SI{10}{mm}$ for the peak displacement due to disturbances by $B'_\mathrm{tr}$.
\begin{figure}[hbt]
	\centering
	\includegraphics{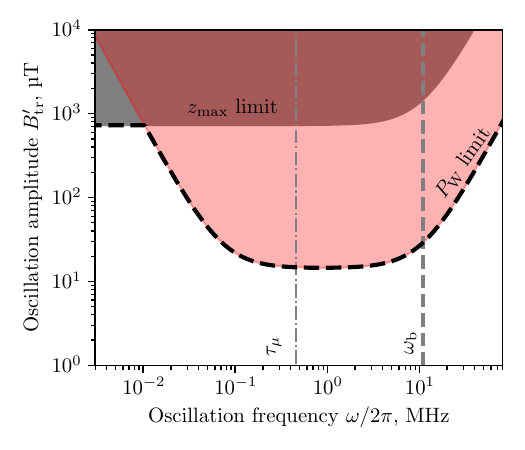}
	\caption{Limit on the amplitude of a transient radial magnetic
		field $B'_\mathrm{tr}$ with oscillation frequency $\omega/2\pi$. For frequencies
		in the range between \SI{200}{\kilo\hertz} and \SI{10}{\mega\hertz} the limit reaches a plateau at
		\SI{14}{\micro\tesla}. The betatron frequency $\omega_\mathrm{b}$ is shown
		with a vertical dashed line, and the rest-frame muon lifetime $\tau_\mu$
		is shown with a dash-dot line. The grey exclusion area is calculated
		using Eq.~\eqref{eq:zmax_external_b} for $z_\mathrm{max} = \SI{50}{mm}$ and
		$z_0 = \SI{40}{mm}$. The thick black dashed line shows the combined
		limits from $z_\mathrm{max}$ and $P_\mathrm{W}$.}
	\label{fig:external_br_limit}
\end{figure}
The limits on the oscillation amplitude of $B'_\mathrm{tr}$ as a function of
its frequency are shown in Fig.~\ref{fig:external_br_limit}.
For frequencies in the range between \SI{200}{\kilo\hertz} and
\SI{10}{\mega\hertz} the limit reaches a plateau at \SI{14}{\micro\tesla}. This
corresponds to longitudinal oscillations of the mean orbit with \SI{0.2}{mm}
amplitude.

The limit presented in Fig.~\ref{fig:external_br_limit} is valid in the case of an external $B$-field oscillating at a single frequency.
In reality, i.e.\ the
residual transient field from the magnetic kick, the signal will contain a spectrum of
frequencies.
One has to then consider the integral over the effect of
$B'_\mathrm{tr}(\omega)$ as in Eq.~\eqref{eq:limit_ext_b_integral}. For
simplicity we take the ratio
\begin{equation}
	\frac{B'_\mathrm{tr}(\omega)}{B'_\mathrm{L}(\omega)} =
	\frac{\dot P_\mathrm{W}(\omega)}{F\dot\Psi},
	\label{eq:limit_ext_b_ratio}
\end{equation}
where $B'_\mathrm{L}(\omega)$ is the external $B$-field that satisfies the
equation $\dot P_\mathrm{W}(\omega) = F\dot \Psi$, and $\dot\Psi$ is the limit
angular velocity of spin precession by an EDM equal to the statistical
sensitivity of the experiment. This ratio can be thought of as what fraction of
the limit angular velocity is induced by an external field $B'_\mathrm{tr}$ at a
given oscillation frequency $\omega$. The integral of the ratio must be less
than 1 to limit the combined effect of a signal with a spectrum of frequencies,
\begin{equation}
	\int_0^\infty \frac{B'_\mathrm{tr}(\omega)}{B'_\mathrm{L}(\omega)}\,d\omega \leq 1.
	\label{eq:limit_ext_b_ratio_integral}
\end{equation}
The relationship $B'_\mathrm{tr}(\omega)$ can be obtained from the inverse
Fourier transform of $B^\Kick_\rho(t)$ (see
Eq.~\eqref{eq:precession_mdm_total_rho}). Its integral after weighing with
$1/B_\mathrm{L}$ gives the fraction of the contribution to the imposed limit of
$\dot \Psi / 4$.



A potential systematic effect associated with a $z$ shift in the average orbit
is a change in the acceptance of the upstream and downstream detectors. While we
anticipate that identification of the $z$ direction of the emitted positrons
will remain unaffected, this aspect nevertheless requires a more comprehensive
investigation. A more detailed discussion of detection-related systematic
effects will be covered in an upcoming publication.

The limits shown in Fig.~\ref{fig:external_br_limit}, result in stringent specification of the magnetic kicker
used to rotate the momentum of injected muons into a stable orbit.
To avoid systematic effects, it is necessary to limit the amplitudes below \SIrange{10}{100}{\micro\tesla} in the frequency band between \SI{30}{kHz} and \SI{30}{MHz}, which will be measured using a laser-based Faraday rotation magnetometer, similar as described in~\cite{Schreckenberger2021}.
Note that even in the case of a too large residual transient magnetic field from the kicker, its
precise knowledge would allow us to correct for the systematic effect.

\section{Conclusions}
In this study, we have presented analytical equations that describe in detail
the precession arising from the AMM in the EM fields
integral to the setup of the proposed muon EDM experiment at PSI\@. Our findings were
verified using \texttt{Geant4} Monte Carlo simulations that utilised realistic field maps generated by Ansys Maxwell.

We identified that the most relevant systematic effects stem from radial magnetic fields that vary with time and a non-zero longitudinal component of the electric field, i.e., parallel to the magnetic field.
The effects of the longitudinal $E$-field can be largely mitigated by employing the CP-violating nature of the EDM by alternating periodically between CW and CCW particle motion in the storage ring.
This can be achieved by switching the polarity of the currents generating the magnetic field, thus inverting the direction of the magnetic field, and correspondingly reversing the injection direction of the muons. The degree of this cancellation depends on the initial
conditions of the experiment -- the muon momentum distribution, the initial polarisation direction, the EM field setup and the overlap between the counter-rotating orbits.

We also provide a qualitative description of the geometric phase, accumulating as a result of spin oscillations in a non-uniform EM field. We found that systematic effects of substantial magnitude can only arise
due to resonances between oscillations around two orthogonal axes, and only if
the relative phase between the two oscillations is non-zero.
One such effect may
originate from the cyclotron motion in the electric field if the axis of this
rotational-symmetric field is displaced and tilted with respect to the muon
orbit's central axis. Geometric phase accumulation due to oscillations
with considerably different periods, such as cyclotron and betatron
oscillations, has negligible impact.


The study presented here is a key contribution to the ongoing effort to search for the muon EDM at PSI. While specific calculations were presented for the initial phase of the experiment, the analytical derivations were kept sufficiently general so as to serve future upgrades aiming for higher sensitivity. The discussed systematic effects could also be relevant for other planned storage-ring EDM searches.

\begin{acknowledgements}
	Several discussions with C. Carli (CERN) are warmly acknowledged.





	This work was supported by the Swiss National Science Fund under grant \textnumero~204118; the European Union's Horizon 2020 research and innovation programme under the Marie Sk\l{}od\-owska-Curie grant agreement \textnumero~884104 (PSI--FELL\-OW--III-3i); the Swiss State Secretariat for Education, Research and Innovation (SERI) under grant \textnumero~MB22.00040; and ETH Zürich under grant \textnumero~ETH-48 18-1.
\end{acknowledgements}

\bibliographystyle{spphys}
\bibliography{muonEDM}

\begin{thebibliography}{10}
\providecommand{\url}[1]{{#1}}
\providecommand{\urlprefix}{URL }
\expandafter\ifx\csname urlstyle\endcsname\relax
  \providecommand{\doi}[1]{DOI \discretionary{}{}{}#1}\else
  \providecommand{\doi}{DOI \discretionary{}{}{}\begingroup
  \urlstyle{rm}\Url}\fi

\bibitem{Pospelov:2013sca}
M.~Pospelov, A.~Ritz, Phys. Rev. D \textbf{89}(5), 056006 (2014).
\newblock \doi{10.1103/PhysRevD.89.056006}

\bibitem{Seng:2014lea}
C.Y. Seng, Phys. Rev. C \textbf{91}(2), 025502 (2015).
\newblock \doi{10.1103/PhysRevC.91.025502}

\bibitem{Abi2021PRL}
B.~Abi, T.~Albahri, S.~Al-Kilani, D.~Allspach, L.P. Alonzi, A.~Anastasi,
  A.~Anisenkov, F.~Azfar, K.~Badgley, S.~Bae\ss{}ler, I.~Bailey, V.A. Baranov,
  E.~Barlas-Yucel, T.~Barrett, E.~Barzi, A.~Basti, F.~Bedeschi, A.~Behnke,
  M.~Berz, M.~Bhattacharya, H.P. Binney, R.~Bjorkquist, P.~Bloom, J.~Bono,
  E.~Bottalico, T.~Bowcock, D.~Boyden, G.~Cantatore, R.M. Carey, J.~Carroll,
  B.C.K. Casey, D.~Cauz, S.~Ceravolo, R.~Chakraborty, S.P. Chang, A.~Chapelain,
  S.~Chappa, S.~Charity, R.~Chislett, J.~Choi, Z.~Chu, T.E. Chupp, M.E.
  Convery, A.~Conway, G.~Corradi, S.~Corrodi, L.~Cotrozzi, J.D. Crnkovic,
  S.~Dabagov, P.M.D. Lurgio, P.T. Debevec, S.D. Falco, P.D. Meo, G.D. Sciascio,
  R.D. Stefano, B.~Drendel, A.~Driutti, V.N. Duginov, M.~Eads, N.~Eggert,
  A.~Epps, J.~Esquivel, M.~Farooq, R.~Fatemi, C.~Ferrari, M.~Fertl, A.~Fiedler,
  A.T. Fienberg, A.~Fioretti, D.~Flay, S.B. Foster, H.~Friedsam, E.~Frle\v{z},
  N.S. Froemming, J.~Fry, C.~Fu, C.~Gabbanini, M.D. Galati, S.~Ganguly,
  A.~Garcia, D.E. Gastler, J.~George, L.K. Gibbons, A.~Gioiosa, K.L.
  Giovanetti, P.~Girotti, W.~Gohn, T.~Gorringe, J.~Grange, S.~Grant, F.~Gray,
  S.~Haciomeroglu, D.~Hahn, T.~Halewood-Leagas, D.~Hampai, F.~Han, E.~Hazen,
  J.~Hempstead, S.~Henry, A.T. Herrod, D.W. Hertzog, G.~Hesketh, A.~Hibbert,
  Z.~Hodge, J.L. Holzbauer, K.W. Hong, R.~Hong, M.~Iacovacci, M.~Incagli,
  C.~Johnstone, J.A. Johnstone, P.~Kammel, M.~Kargiantoulakis, M.~Karuza,
  J.~Kaspar, D.~Kawall, L.~Kelton, A.~Keshavarzi, D.~Kessler, K.S. Khaw,
  Z.~Khechadoorian, N.V. Khomutov, B.~Kiburg, M.~Kiburg, O.~Kim, S.C. Kim, Y.I.
  Kim, B.~King, N.~Kinnaird, M.~Korostelev, I.~Kourbanis, E.~Kraegeloh, V.A.
  Krylov, A.~Kuchibhotla, N.A. Kuchinskiy, K.R. Labe, J.~LaBounty,
  M.~Lancaster, M.J. Lee, S.~Lee, S.~Leo, B.~Li, D.~Li, L.~Li, I.~Logashenko,
  A.L. Campos, A.~Luc\`a, G.~Lukicov, G.~Luo, A.~Lusiani, A.L. Lyon, B.~MacCoy,
  R.~Madrak, K.~Makino, F.~Marignetti, S.~Mastroianni, S.~Maxfield, M.~McEvoy,
  W.~Merritt, A.A. Mikhailichenko, J.P. Miller, S.~Miozzi, J.P. Morgan, W.M.
  Morse, J.~Mott, E.~Motuk, A.~Nath, D.~Newton, H.~Nguyen, M.~Oberling,
  R.~Osofsky, J.F. Ostiguy, S.~Park, G.~Pauletta, G.M. Piacentino, R.N. Pilato,
  K.T. Pitts, B.~Plaster, D.~Po\v{c}, N.~Pohlman, C.C. Polly, M.~Popovic,
  J.~Price, B.~Quinn, N.~Raha, S.~Ramachandran, E.~Ramberg, N.T. Rider, J.L.
  Ritchie, B.L. Roberts, D.L. Rubin, L.~Santi, D.~Sathyan, H.~Schellman,
  C.~Schlesier, A.~Schreckenberger, Y.K. Semertzidis, Y.M. Shatunov,
  D.~Shemyakin, M.~Shenk, D.~Sim, M.W. Smith, A.~Smith, A.K. Soha, M.~Sorbara,
  D.~St{\"o}ckinger, J.~Stapleton, D.~Still, C.~Stoughton, D.~Stratakis,
  C.~Strohman, T.~Stuttard, H.E. Swanson, G.~Sweetmore, D.A. Sweigart, M.J.
  Syphers, D.A. Tarazona, T.~Teubner, A.E. Tewsley-Booth, K.~Thomson,
  V.~Tishchenko, N.H. Tran, W.~Turner, E.~Valetov, D.~Vasilkova, G.~Venanzoni,
  V.P. Volnykh, T.~Walton, M.~Warren, A.~Weisskopf, L.~Welty-Rieger,
  M.~Whitley, P.~Winter, A.~Wolski, M.~Wormald, W.~Wu, C.~Yoshikawa, Phys. Rev.
  Lett. \textbf{126}, 141801 (2021).
\newblock \doi{10.1103/PhysRevLett.126.141801}.
\newblock
  \urlprefix\url{https://link.aps.org/doi/10.1103/PhysRevLett.126.141801}

\bibitem{FNAL2023PRL}
D.P. Aguillard, T.~Albahri, D.~Allspach, A.~Anisenkov, K.~Badgley,
  S.~Bae\ss{}ler, I.~Bailey, L.~Bailey, V.A. Baranov, E.~Barlas-Yucel,
  T.~Barrett, E.~Barzi, F.~Bedeschi, M.~Berz, M.~Bhattacharya, H.P. Binney,
  P.~Bloom, J.~Bono, E.~Bottalico, T.~Bowcock, S.~Braun, M.~Bressler,
  G.~Cantatore, R.M. Carey, B.C.K. Casey, D.~Cauz, R.~Chakraborty,
  A.~Chapelain, S.~Chappa, S.~Charity, C.~Chen, M.~Cheng, R.~Chislett, Z.~Chu,
  T.E. Chupp, C.~Claessens, M.E. Convery, S.~Corrodi, L.~Cotrozzi, J.D.
  Crnkovic, S.~Dabagov, P.T. Debevec, S.~Di~Falco, G.~Di~Sciascio, B.~Drendel,
  A.~Driutti, V.N. Duginov, M.~Eads, A.~Edmonds, J.~Esquivel, M.~Farooq,
  R.~Fatemi, C.~Ferrari, M.~Fertl, A.T. Fienberg, A.~Fioretti, D.~Flay, S.B.
  Foster, H.~Friedsam, N.S. Froemming, C.~Gabbanini, I.~Gaines, M.D. Galati,
  S.~Ganguly, A.~Garcia, J.~George, L.K. Gibbons, A.~Gioiosa, K.L. Giovanetti,
  P.~Girotti, W.~Gohn, L.~Goodenough, T.~Gorringe, J.~Grange, S.~Grant,
  F.~Gray, S.~Haciomeroglu, T.~Halewood-Leagas, D.~Hampai, F.~Han,
  J.~Hempstead, D.W. Hertzog, G.~Hesketh, E.~Hess, A.~Hibbert, Z.~Hodge, K.W.
  Hong, R.~Hong, T.~Hu, Y.~Hu, M.~Iacovacci, M.~Incagli, P.~Kammel,
  M.~Kargiantoulakis, M.~Karuza, J.~Kaspar, D.~Kawall, L.~Kelton,
  A.~Keshavarzi, D.S. Kessler, K.S. Khaw, Z.~Khechadoorian, N.V. Khomutov,
  B.~Kiburg, M.~Kiburg, O.~Kim, N.~Kinnaird, E.~Kraegeloh, V.A. Krylov, N.A.
  Kuchinskiy, K.R. Labe, J.~LaBounty, M.~Lancaster, S.~Lee, B.~Li, D.~Li,
  L.~Li, I.~Logashenko, A.~Lorente~Campos, Z.~Lu, A.~Luc\`a, G.~Lukicov,
  A.~Lusiani, A.L. Lyon, B.~MacCoy, R.~Madrak, K.~Makino, S.~Mastroianni, J.P.
  Miller, S.~Miozzi, B.~Mitra, J.P. Morgan, W.M. Morse, J.~Mott, A.~Nath, J.K.
  Ng, H.~Nguyen, Y.~Oksuzian, Z.~Omarov, R.~Osofsky, S.~Park, G.~Pauletta, G.M.
  Piacentino, R.N. Pilato, K.T. Pitts, B.~Plaster, D.~Po\ifmmode \check{c}\else
  \v{c}\fi{}ani\ifmmode~\acute{c}\else \'{c}\fi{}, N.~Pohlman, C.C. Polly,
  J.~Price, B.~Quinn, M.U.H. Qureshi, S.~Ramachandran, E.~Ramberg, R.~Reimann,
  B.L. Roberts, D.L. Rubin, L.~Santi, C.~Schlesier, A.~Schreckenberger, Y.K.
  Semertzidis, D.~Shemyakin, M.~Sorbara, D.~St\"ockinger, J.~Stapleton,
  D.~Still, C.~Stoughton, D.~Stratakis, H.E. Swanson, G.~Sweetmore, D.A.
  Sweigart, M.J. Syphers, D.A. Tarazona, T.~Teubner, A.E. Tewsley-Booth,
  V.~Tishchenko, N.H. Tran, W.~Turner, E.~Valetov, D.~Vasilkova, G.~Venanzoni,
  V.P. Volnykh, T.~Walton, A.~Weisskopf, L.~Welty-Rieger, P.~Winter, Y.~Wu,
  B.~Yu, M.~Yucel, Y.~Zeng, C.~Zhang, Phys. Rev. Lett. \textbf{131}, 161802
  (2023).
\newblock \doi{10.1103/PhysRevLett.131.161802}.
\newblock
  \urlprefix\url{https://link.aps.org/doi/10.1103/PhysRevLett.131.161802}

\bibitem{Aoyama:2020ynm}
T.~Aoyama, N.~Asmussen, M.~Benayoun, J.~Bijnens, T.~Blum, M.~Bruno, I.~Caprini,
  C.M.C. Calame, M.~C{\`e}, G.~Colangelo, F.~Curciarello, H.~Czy{\.z},
  I.~Danilkin, M.~Davier, C.T.H. Davies, M.D. Morte, S.I. Eidelman, A.X.
  El-Khadra, A.~G{\'e}rardin, D.~Giusti, M.~Golterman, S.~Gottlieb,
  V.~G{\"u}lpers, F.~Hagelstein, M.~Hayakawa, G.~Herdo{\'\i}za, D.W. Hertzog,
  A.~Hoecker, M.~Hoferichter, B.L. Hoid, R.J. Hudspith, F.~Ignatov,
  T.~Izubuchi, F.~Jegerlehner, L.~Jin, A.~Keshavarzi, T.~Kinoshita, B.~Kubis,
  A.~Kupich, A.~Kup{\'s}{\'c}, L.~Laub, C.~Lehner, L.~Lellouch, I.~Logashenko,
  B.~Malaescu, K.~Maltman, M.K. Marinkovi{\'c}, P.~Masjuan, A.S. Meyer, H.B.
  Meyer, T.~Mibe, K.~Miura, S.E. M{\"u}ller, M.~Nio, D.~Nomura, A.~Nyffeler,
  V.~Pascalutsa, M.~Passera, E.P. del Rio, S.~Peris, A.~Portelli, M.~Procura,
  C.F. Redmer, B.L. Roberts, P.~S{\'a}nchez-Puertas, S.~Serednyakov,
  B.~Shwartz, S.~Simula, D.~St{\"o}ckinger, H.~St{\"o}ckinger-Kim, P.~Stoffer,
  T.~Teubner, R.V. de~Water, M.~Vanderhaeghen, G.~Venanzoni, G.~von Hippel,
  H.~Wittig, Z.~Zhang, M.N. Achasov, A.~Bashir, N.~Cardoso, B.~Chakraborty,
  E.H. Chao, J.~Charles, A.~Crivellin, O.~Deineka, A.~Denig, C.~DeTar, C.A.
  Dominguez, A.E. Dorokhov, V.P. Druzhinin, G.~Eichmann, M.~Fael, C.S. Fischer,
  E.~G{\'a}miz, Z.~Gelzer, J.R. Green, S.~Guellati-Khelifa, D.~Hatton,
  N.~Hermansson-Truedsson, S.~Holz, B.~H{\"o}rz, M.~Knecht, J.~Koponen, A.S.
  Kronfeld, J.~Laiho, S.~Leupold, P.B. Mackenzie, W.J. Marciano, C.~McNeile,
  D.~Mohler, J.~Monnard, E.T. Neil, A.V. Nesterenko, K.~Ottnad, V.~Pauk, A.E.
  Radzhabov, E.~de~Rafael, K.~Raya, A.~Risch, A.~Rodr{\'\i}guez-S{\'a}nchez,
  P.~Roig, T.S. Jos{\'e}, E.P. Solodov, R.~Sugar, K.Y. Todyshev, A.~Vainshtein,
  A.V. Avil{\'e}s-Casco, E.~Weil, J.~Wilhelm, R.~Williams, A.S. Zhevlakov,
  Phys. Rep. \textbf{887}, 1 (2020).
\newblock \doi{10.1016/j.physrep.2020.07.006}

\bibitem{Farley2004PRL}
F.J.M. Farley, K.~Jungmann, J.P. Miller, W.M. Morse, Y.F. Orlov, B.L. Roberts,
  Y.K. Semertzidis, A.~Silenko, E.J. Stephenson, Phys. Rev. Lett. \textbf{93},
  052001 (2004).
\newblock \doi{10.1103/PhysRevLett.93.052001}

\bibitem{Khriplovich1998}
I.B. Khriplovich, Physics Letters B \textbf{444}(1-2), 98 (1998).
\newblock \doi{10.1016/s0370-2693(98)01353-7}.
\newblock \urlprefix\url{https://doi.org/10.1016/s0370-2693(98)01353-7}

\bibitem{Semertzidis2001}
Y.K. Semertzidis, in \emph{{AIP} Conference Proceedings} ({AIP}, 2001).
\newblock \doi{10.1063/1.1374993}.
\newblock \urlprefix\url{https://doi.org/10.1063/1.1374993}

\bibitem{Agostinelli2003}
S.~Agostinelli, J.~Allison, K.~Amako, J.~Apostolakis, H.~Araujo, P.~Arce,
  M.~Asai, D.~Axen, S.~Banerjee, G.~Barrand, F.~Behner, L.~Bellagamba,
  J.~Boudreau, L.~Broglia, A.~Brunengo, H.~Burkhardt, S.~Chauvie, J.~Chuma,
  R.~Chytracek, G.~Cooperman, G.~Cosmo, P.~Degtyarenko,
  A.~Dell{\textquotesingle}Acqua, G.~Depaola, D.~Dietrich, R.~Enami,
  A.~Feliciello, C.~Ferguson, H.~Fesefeldt, G.~Folger, F.~Foppiano, A.~Forti,
  S.~Garelli, S.~Giani, R.~Giannitrapani, D.~Gibin, J.J.G. Cadenas,
  I.~Gonz{\'{a}}lez, G.G. Abril, G.~Greeniaus, W.~Greiner, V.~Grichine,
  A.~Grossheim, S.~Guatelli, P.~Gumplinger, R.~Hamatsu, K.~Hashimoto, H.~Hasui,
  A.~Heikkinen, A.~Howard, V.~Ivanchenko, A.~Johnson, F.W. Jones,
  J.~Kallenbach, N.~Kanaya, M.~Kawabata, Y.~Kawabata, M.~Kawaguti, S.~Kelner,
  P.~Kent, A.~Kimura, T.~Kodama, R.~Kokoulin, M.~Kossov, H.~Kurashige,
  E.~Lamanna, T.~Lamp{\'{e}}n, V.~Lara, V.~Lefebure, F.~Lei, M.~Liendl,
  W.~Lockman, F.~Longo, S.~Magni, M.~Maire, E.~Medernach, K.~Minamimoto, P.M.
  de~Freitas, Y.~Morita, K.~Murakami, M.~Nagamatu, R.~Nartallo, P.~Nieminen,
  T.~Nishimura, K.~Ohtsubo, M.~Okamura, S.~O{\textquotesingle}Neale, Y.~Oohata,
  K.~Paech, J.~Perl, A.~Pfeiffer, M.G. Pia, F.~Ranjard, A.~Rybin, S.~Sadilov,
  E.D. Salvo, G.~Santin, T.~Sasaki, N.~Savvas, Y.~Sawada, S.~Scherer, S.~Sei,
  V.~Sirotenko, D.~Smith, N.~Starkov, H.~Stoecker, J.~Sulkimo, M.~Takahata,
  S.~Tanaka, E.~Tcherniaev, E.S. Tehrani, M.~Tropeano, P.~Truscott, H.~Uno,
  L.~Urban, P.~Urban, M.~Verderi, A.~Walkden, W.~Wander, H.~Weber, J.P.
  Wellisch, T.~Wenaus, D.C. Williams, D.~Wright, T.~Yamada, H.~Yoshida,
  D.~Zschiesche, Nucl. Instrum. Meth. A \textbf{506}(3), 250 (2003).
\newblock \doi{10.1016/s0168-9002(03)01368-8}.
\newblock \urlprefix\url{https://doi.org/10.1016/s0168-9002(03)01368-8}

\bibitem{Allison2006}
J.~Allison, K.~Amako, J.~Apostolakis, H.~Araujo, P.A. Dubois, M.~Asai,
  G.~Barrand, R.~Capra, S.~Chauvie, R.~Chytracek, G.A.P. Cirrone, G.~Cooperman,
  G.~Cosmo, G.~Cuttone, G.G. Daquino, M.~Donszelmann, M.~Dressel, G.~Folger,
  F.~Foppiano, J.~Generowicz, V.~Grichine, S.~Guatelli, P.~Gumplinger,
  A.~Heikkinen, I.~Hrivnacova, A.~Howard, S.~Incerti, V.~Ivanchenko,
  T.~Johnson, F.~Jones, T.~Koi, R.~Kokoulin, M.~Kossov, H.~Kurashige, V.~Lara,
  S.~Larsson, F.~Lei, O.~Link, F.~Longo, M.~Maire, A.~Mantero, B.~Mascialino,
  I.~McLaren, P.M. Lorenzo, K.~Minamimoto, K.~Murakami, P.~Nieminen,
  L.~Pandola, S.~Parlati, L.~Peralta, J.~Perl, A.~Pfeiffer, M.G. Pia, A.~Ribon,
  P.~Rodrigues, G.~Russo, S.~Sadilov, G.~Santin, T.~Sasaki, D.~Smith,
  N.~Starkov, S.~Tanaka, E.~Tcherniaev, B.~Tome, A.~Trindade, P.~Truscott,
  L.~Urban, M.~Verderi, A.~Walkden, J.P. Wellisch, D.C. Williams, D.~Wright,
  H.~Yoshida, {IEEE} Transactions on Nuclear Science \textbf{53}(1), 270
  (2006).
\newblock \doi{10.1109/tns.2006.869826}.
\newblock \urlprefix\url{https://doi.org/10.1109/tns.2006.869826}

\bibitem{Allison2016}
J.~Allison, K.~Amako, J.~Apostolakis, P.~Arce, M.~Asai, T.~Aso, E.~Bagli,
  A.~Bagulya, S.~Banerjee, G.~Barrand, B.R. Beck, A.G. Bogdanov, D.~Brandt,
  J.M.C. Brown, H.~Burkhardt, P.~Canal, D.~Cano-Ott, S.~Chauvie, K.~Cho, G.A.P.
  Cirrone, G.~Cooperman, M.A. Cort{\'{e}}s-Giraldo, G.~Cosmo, G.~Cuttone,
  G.~Depaola, L.~Desorgher, X.~Dong, A.~Dotti, V.D. Elvira, G.~Folger,
  Z.~Francis, A.~Galoyan, L.~Garnier, M.~Gayer, K.L. Genser, V.M. Grichine,
  S.~Guatelli, P.~Gu{\`{e}}ye, P.~Gumplinger, A.S. Howard,
  I.~H{\v{r}}ivn{\'{a}}{\v{c}}ov{\'{a}}, S.~Hwang, S.~Incerti, A.~Ivanchenko,
  V.N. Ivanchenko, F.W. Jones, S.Y. Jun, P.~Kaitaniemi, N.~Karakatsanis,
  M.~Karamitros, M.~Kelsey, A.~Kimura, T.~Koi, H.~Kurashige, A.~Lechner, S.B.
  Lee, F.~Longo, M.~Maire, D.~Mancusi, A.~Mantero, E.~Mendoza, B.~Morgan,
  K.~Murakami, T.~Nikitina, L.~Pandola, P.~Paprocki, J.~Perl,
  I.~Petrovi{\'{c}}, M.G. Pia, W.~Pokorski, J.M. Quesada, M.~Raine, M.A. Reis,
  A.~Ribon, A.R. Fira, F.~Romano, G.~Russo, G.~Santin, T.~Sasaki, D.~Sawkey,
  J.I. Shin, I.I. Strakovsky, A.~Taborda, S.~Tanaka, B.~Tom{\'{e}}, T.~Toshito,
  H.N. Tran, P.R. Truscott, L.~Urban, V.~Uzhinsky, J.M. Verbeke, M.~Verderi,
  B.L. Wendt, H.~Wenzel, D.H. Wright, D.M. Wright, T.~Yamashita, J.~Yarba,
  H.~Yoshida, Nuclear Instruments and Methods in Physics Research Section A:
  Accelerators, Spectrometers, Detectors and Associated Equipment \textbf{835},
  186 (2016).
\newblock \doi{10.1016/j.nima.2016.06.125}.
\newblock \urlprefix\url{https://doi.org/10.1016/j.nima.2016.06.125}

\bibitem{Adelmann2010JPG}
A.~Adelmann, K.~Kirch, C.J.G. Onderwater, T.~Schietinger, J. Phys. G
  \textbf{37}, 085001 (2010).
\newblock \doi{10.1088/0954-3899/37/8/085001}

\bibitem{Adelmann2021arXiv}
A.~Adelmann, M.~Backhaus, C.C. Barajas, N.~Berger, T.~Bowcock, C.~Calzolaio,
  G.~Cavoto, R.~Chislett, A.~Crivellin, M.~Daum, M.~Fertl, M.~Giovannozzi,
  G.~Hesketh, M.~Hildebrandt, I.~Keshelashvili, A.~Keshavarzi, K.S. Khaw,
  K.~Kirch, A.~Kozlinskiy, A.~Knecht, M.~Lancaster, B.~M{\"{a}}rkisch, F.M.
  Aeschbacher, F.~Méot, A.~Nass, A.~Papa, J.~Pretz, J.~Price, F.~Rathmann,
  F.~Renga, M.~Sakurai, P.~Schmidt-Wellenburg, A.~Sch{\"{o}}ning, M.~Schott,
  C.~Voena, J.~Vossebeld, F.~Wauters, P.~Winter.
\newblock Search for a muon {EDM} using the frozen-spin technique (2021).
\newblock \doi{10.48550/ARXIV.2102.08838}.
\newblock \urlprefix\url{https://arxiv.org/abs/2102.08838}

\bibitem{Barna2017}
D.~Barna, Phys. Rev. Acc. Beams \textbf{20}(4), 041002 (2017).
\newblock \doi{10.1103/PhysRevAccelBeams.20.041002}

\bibitem{Iinuma2016}
H.~Iinuma, H.~Nakayama, K.~Oide, K.~Sasaki, N.~Saito, T.~Mibe, M.~Abe, Nucl.
  Instrum. Meth. A \textbf{832}, 51 (2016).
\newblock \doi{10.1016/j.nima.2016.05.126}.
\newblock \urlprefix\url{https://doi.org/10.1016/j.nima.2016.05.126}

\bibitem{Bennett2009PRD}
G.W. Bennett, B.~Bousquet, H.N. Brown, G.~Bunce, R.M. Carey, P.~Cushman, G.T.
  Danby, P.T. Debevec, M.~Deile, H.~Deng, W.~Deninger, S.K. Dhawan, V.P.
  Druzhinin, L.~Duong, E.~Efstathiadis, F.J.M. Farley, G.V. Fedotovich,
  S.~Giron, F.E. Gray, D.~Grigoriev, M.~Grosse-Perdekamp, A.~Grossmann, M.F.
  Hare, D.W. Hertzog, X.~Huang, V.W. Hughes, M.~Iwasaki, K.~Jungmann,
  D.~Kawall, M.~Kawamura, B.I. Khazin, J.~Kindem, F.~Krienen, I.~Kronkvist,
  A.~Lam, R.~Larsen, Y.Y. Lee, I.~Logashenko, R.~McNabb, W.~Meng, J.~Mi, J.P.
  Miller, Y.~Mizumachi, W.M. Morse, D.~Nikas, C.J.G. Onderwater, Y.~Orlov, C.S.
  {\"O}zben, J.M. Paley, Q.~Peng, C.C. Polly, J.~Pretz, R.~Prigl, G.Z. Putlitz,
  T.~Qian, S.I. Redin, O.~Rind, B.L. Roberts, N.~Ryskulov, S.~Sedykh, Y.K.
  Semertzidis, P.~Shagin, Y.M. Shatunov, E.P. Sichtermann, E.~Solodov,
  M.~Sossong, A.~Steinmetz, L.R. Sulak, C.~Timmermans, A.~Trofimov, D.~Urner,
  P.~von Walter, D.~Warburton, D.~Winn, A.~Yamamoto, D.~Zimmerman, Phys. Rev. D
  \textbf{80}(5), 052008 (2009).
\newblock \doi{10.1103/PhysRevD.80.052008}.
\newblock \urlprefix\url{https://link.aps.org/doi/10.1103/PhysRevD.80.052008}

\bibitem{Thomas1926}
L.H. Thomas, Nature \textbf{117}(2945), 514 (1926).
\newblock \doi{10.1038/117514a0}.
\newblock \urlprefix\url{https://doi.org/10.1038/117514a0}

\bibitem{Thomas1927}
L.H. Thomas, Lond. Edinb. Dublin philos. mag. j. sci. \textbf{3}(13), 1 (1927).
\newblock \doi{10.1080/14786440108564170}.
\newblock \urlprefix\url{https://doi.org/10.1080/14786440108564170}

\bibitem{PhysRevLett.2.435}
V.~Bargmann, L.~Michel, V.L. Telegdi, Phys. Rev. Lett. \textbf{2}, 435 (1959).
\newblock \doi{10.1103/PhysRevLett.2.435}.
\newblock \urlprefix\url{https://link.aps.org/doi/10.1103/PhysRevLett.2.435}

\bibitem{HajTahar2021}
M.H. Tahar, C.~Carli, Phys. Rev. Accel. Beams \textbf{24}, 034003 (2021).
\newblock \doi{10.1103/PhysRevAccelBeams.24.034003}.
\newblock
  \urlprefix\url{https://link.aps.org/doi/10.1103/PhysRevAccelBeams.24.034003}

\bibitem{Rabi1954}
I.I. Rabi, N.F. Ramsey, J.~Schwinger, Rev. Mod. Phys. \textbf{26}(2), 167
  (1954).
\newblock \doi{10.1103/RevModPhys.26.167}

\bibitem{Berry1984}
M.V. Berry, Proc. Roy. Soc. Lond. \textbf{A392}, 45 (1984).
\newblock \doi{10.1098/rspa.1984.0023}

\bibitem{Ansys_ED}
Ansoft, \emph{Ansys\textsuperscript{\textregistered} Academic Research
  Electronics Desktop} (2022).
\newblock Ver. 2022R1

\bibitem{Hannay:1985}
J.H. Hannay, J. Phys. A: Math. Gen. \textbf{18}, 221 (1985).
\newblock \doi{10.1088/0305-4470/18/2/011}.
\newblock \urlprefix\url{https://doi.org/10.1088/0305-4470/18/2/011}

\bibitem{Sakurai2011}
J.J. Sakurai, J.~Napolitano, \emph{{Modern quantum mechanics; 2nd ed.}}
  (Addison-Wesley, San Francisco, CA, 2011).
\newblock \urlprefix\url{https://cds.cern.ch/record/1341875}

\bibitem{Carli2021}
C.~Carli, M.H. Tahar, Proceedings of the 12th International Particle
  Accelerator Conference \textbf{IPAC2021}, Brazil (2021).
\newblock \doi{10.18429/JACOW-IPAC2021-MOPAB178}.
\newblock
  \urlprefix\url{https://jacow.org/ipac2021/doi/JACoW-IPAC2021-MOPAB178.html}

\bibitem{Carli2022}
C.~Carli, M.H. Tahar, Physical Review Accelerators and Beams \textbf{25}(6)
  (2022).
\newblock \doi{10.1103/physrevaccelbeams.25.064001}.
\newblock \urlprefix\url{https://doi.org/10.1103/physrevaccelbeams.25.064001}

\bibitem{Lszl2018}
A.~L{\'{a}}szl{\'{o}}, Z.~Zimbor{\'{a}}s, Classical and Quantum Gravity
  \textbf{35}(17), 175003 (2018).
\newblock \doi{10.1088/1361-6382/aacfee}.
\newblock \urlprefix\url{https://doi.org/10.1088/1361-6382/aacfee}

\bibitem{Pretz2020}
J.~Pretz, Journal of Physics: Conference Series \textbf{1586}(1), 012043
  (2020).
\newblock \doi{10.1088/1742-6596/1586/1/012043}.
\newblock \urlprefix\url{https://doi.org/10.1088/1742-6596/1586/1/012043}

\bibitem{Kobach2016}
A.~Kobach, Nuclear Physics B \textbf{911}, 206 (2016).
\newblock \doi{10.1016/j.nuclphysb.2016.08.011}.
\newblock \urlprefix\url{https://doi.org/10.1016/j.nuclphysb.2016.08.011}

\bibitem{Jackson1976}
J.D. Jackson, Reviews of Modern Physics \textbf{48}(3), 417 (1976).
\newblock \doi{10.1103/revmodphys.48.417}.
\newblock \urlprefix\url{https://doi.org/10.1103/revmodphys.48.417}

\bibitem{Brown1982}
L.S. Brown, G.~Gabrielse, Physical Review A \textbf{25}(4), 2423 (1982).
\newblock \doi{10.1103/physreva.25.2423}.
\newblock \urlprefix\url{https://doi.org/10.1103/physreva.25.2423}

\bibitem{Schreckenberger2021}
A.P. Schreckenberger, D.~Allspach, D.~Barak, J.~Bohn, C.~Bradford, D.~Cauz,
  S.P. Chang, A.~Chapelain, S.~Chappa, S.~Charity, R.~Chislett, J.~Esquivel,
  C.~Ferrari, A.~Fioretti, C.~Gabbanini, M.D. Galati, L.~Gibbons, J.L.
  Holzbauer, M.~Incagli, C.~Jensen, J.~Kaspar, D.~Kawall, A.~Keshavarzi, D.S.
  Kessler, B.~Kiburg, G.~Krafczyk, R.~Madrak, A.A. Mikhailichenko, H.~Nguyen,
  K.~Overhage, S.~Park, H.~Pfeffer, C.C. Polly, M.~Popovic, R.~Rivera, B.L.
  Roberts, D.~Rubin, Y.K. Semertzidis, J.~Stapleton, C.~Stoughton, E.~Voirin,
  D.~Wolff, Nuclear Instruments and Methods in Physics Research Section A:
  Accelerators, Spectrometers, Detectors and Associated Equipment
  \textbf{1011}, 165597 (2021).
\newblock \doi{10.1016/j.nima.2021.165597}.
\newblock \urlprefix\url{https://doi.org/10.1016/j.nima.2021.165597}

\bibitem{Catmull1974}
E.~Catmull, R.~Rom, in \emph{Computer Aided Geometric Design} (Elsevier, 1974),
  pp. 317--326.
\newblock \doi{10.1016/b978-0-12-079050-0.50020-5}.
\newblock \urlprefix\url{https://doi.org/10.1016/b978-0-12-079050-0.50020-5}

\bibitem{NASAloop}
J.C. Simpson, J.E. Lane, C.D. Immer, R.C. Youngquist, T.~Steinrock, Preprint
  (Draft being sent to journal)  (2001).
\newblock \urlprefix\url{https://ntrs.nasa.gov/search.jsp?R=20010038494}.
\newblock {NASA} Kennedy Space Center, Cocoa Beach, FL, United States

\end{thebibliography}

\clearpage
\appendix
\section{Verification using Geant4 spin tracking}\label{sec:g4_verification}
\subsection{General considerations}
To verify the analytical equations, a model of the experimental setup proposed was
developed using the \texttt{Geant4} Monte Carlo simulation toolkit.
The EM fields of the experiment can be calculated analytically or
interpolated from field maps. The field maps are generated by the \texttt{ANSYS Maxwell3D}
FEM software.
The simulation has three major EM field components:
\begin{enumerate}
	\item The main solenoid magnetic field, with a constant value along $z$, or a
	      field map supplied by FEM simulations.
	\item Radial electric field given by Eq.~\eqref{eq:e_field_rotated} with the option to add a constant or uniform component in the $z$ direction or FEM simulated field map.
	\item Weakly-focusing field modelled in ANSYS as a single circular coil with radius $R = \SI{65}{mm}$.
\end{enumerate}

Muons start with zero momentum in the $z$ direction, since this is the
initial condition for a stored muon. The simulation tracks the spin orientation
in the reference frame of the muon and records it as a function of time. It can
also track the direction with respect to a reference frame defined by the
experimental setup, e.g. the solenoid magnet.

In terms of simulation requirements, the experimental setup of the muon EDM experiment is
situated at the frontier between accelerator/storage ring physics and detector
physics. The storage ring is fully contained within a single solenoid, and the
beam focusing is performed by a circular coil. To maximise sensitivity, all positron detectors are
located as close as possible to the stored muons. Thus, the use of
the \texttt{Geant4} toolkit is highly motivated as it provides an easy
implementation of complex electromagnetic fields together with the detector geometries.
However, its tracking (equation of motion integrator) is mainly developed for
the purposes of single-pass systems. In the case of the muon EDM experiment, the
muons travel a significant distance within the storage ring before decaying. The
Phase~I experiment muons perform 880 turns per muon lifetime, equivalent to about
\SI{180}{m}. Therefore, before proceeding with the verification of the analytical
equations, the capabilities of the \texttt{Geant4} simulation for accurate
tracking and interpolation of EM field grids were tested.

\subsection{Verification of tracking and
	interpolation}\label{sec:tracking_verification}
\begin{figure*}[htb]
	\centering
	\includegraphics[width=0.49\linewidth]{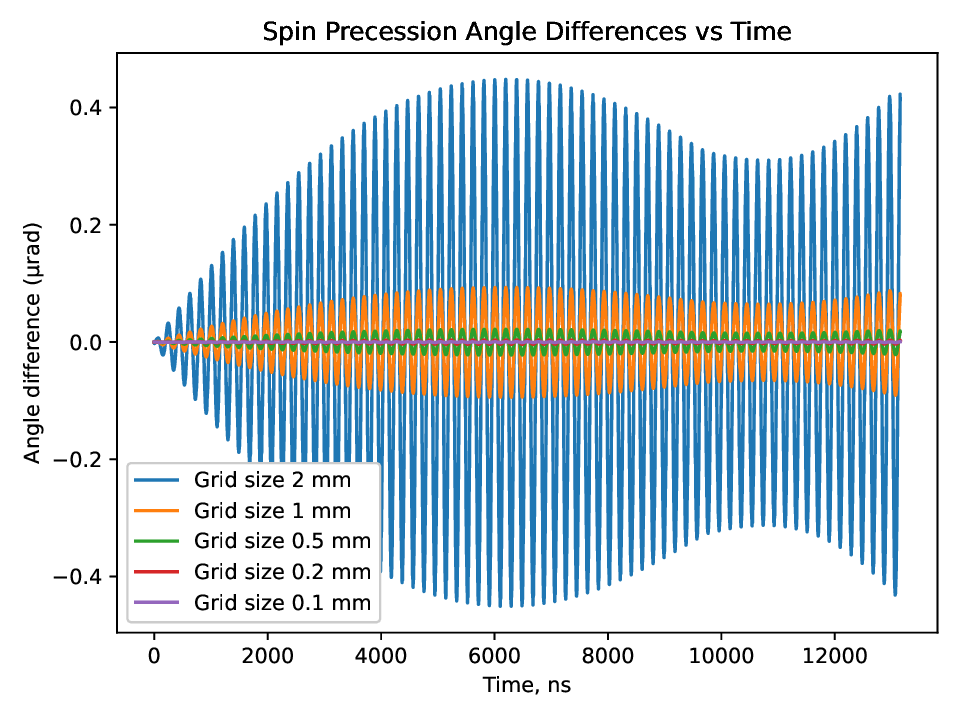}
	\includegraphics[width=0.49\linewidth]{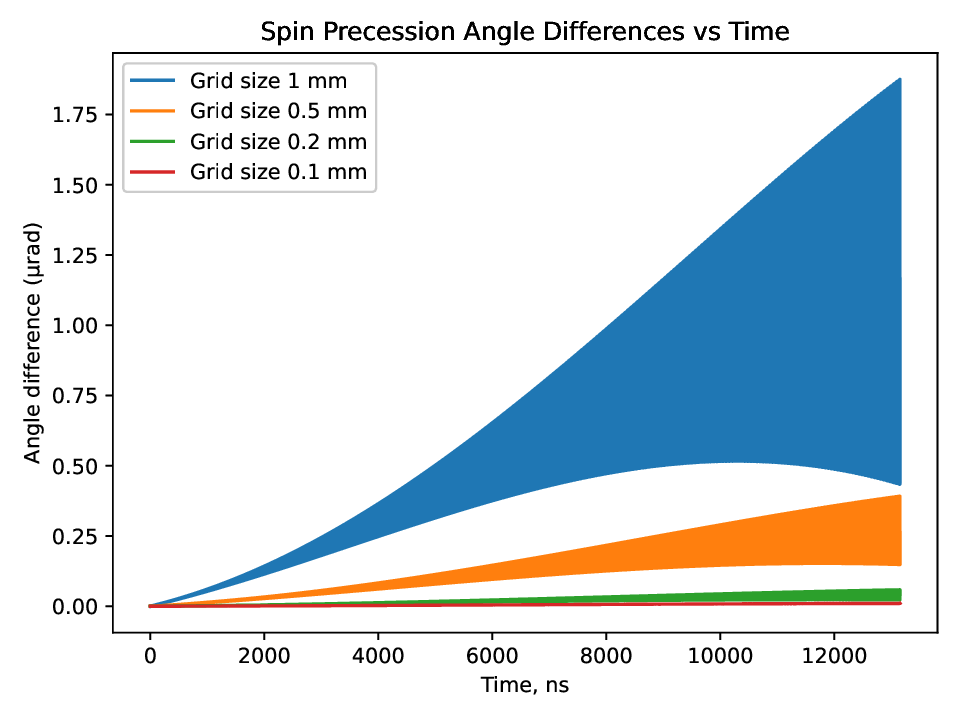}
	\caption{Time evolution of the spin angle difference as a function of the grid size for the $B$-field
		(left) or $E$-field (right). The expected signal should be in the order of \SI{21}{\micro rad/\micro s} and
		\SI{1.3}{\micro rad/\micro s} for the Phase~I and Phase~II experiment, respectively. $E$-field grid size of \SI{0.2}{mm} and $B$-field grid size
		of \SI{0.5}{mm} ensure sufficiently low deviations while still maintaining reasonable
		map size. Note that the curves appear like uniformly filled shapes due to very high frequency cyclotron oscillations.}
	\label{fig:interpolation_benchmark}
\end{figure*}

To track the muons inside the storage ring volume, the Monte Carlo code requires
the values of the EM fields at arbitrary positions in space, whereas all
practically useful methods can only provide those on a grid. We have implemented
a Catmull-Rom cubic spline~\cite{Catmull1974} to interpolate the field at a
point from a regular 3-dimensional field map. To benchmark the performance of
the interpolator, we use an exact analytical solution of the $B$-field of a
circular current loop~\cite{NASAloop} and calculate the field on a regular grid
with step sizes from \SI{0.1}{mm} to \SI{2}{mm}. The numerical simulations were then launched
with a \SI{28}{MeV/\textit{c}} muon starting from a fixed position, which was then tracked
through the field once using its exact analytical expression or by interpolating it
from the regular grid field maps. The difference in spin phase out of the
orbit plane $\Psi$ was calculated between the simulations using the interpolated field, or the exact
solution. The results are shown in
Fig.~\ref{fig:interpolation_benchmark} (left, for the $B$-field, right, for the $E$-field cases). All grid sizes tested below \SI{2}{mm}
result in an acceptable $B$-field approximation in relation to the expected Phase~I
or even Phase~II target sensitivities, \SI{21}{\micro rad/\micro s} and
\SI{1.3}{\micro rad/\micro s}, respectively. The same study was done for the
$E$-field interpolation by generating field maps with grid size from \SI{0.1}{mm} to \SI{1}{mm} from the exact analytical
solution for coaxial cylindrical electrodes. As the spin of muons at that
momentum is more sensitive to the $E$-field, the necessary grid spacing that
ensures a good approximation is \SI{0.2}{mm} or lower.

The sixth-order Dormand-Prince integration routine of
\texttt{Geant4}~\cite{Agostinelli2003} was used for the tracking of muon position and spin throughout the simulations. To determine the optimal step size of the
integrator, we performed tracking simulations with step sizes of fractions of
$\sqrt{2}$ from \SI{0.014}{mm} to \SI{2.82}{mm}. The irrational numbers for the
step size were chosen so as to avoid effects due to resonances between the integration
step and field map grids. The smallest step size generates more than $10^4$
integration steps per rotation and was used as a reference. For all
integration step sizes studied, the deviation of the spin phase $\Psi$ from the
reference trajectory was below \SI{0.1}{\nano rad/\micro s} or four orders of
magnitude below the signal at the statistical sensitivity of the Phase~II
experiment. However, a step size of $\sqrt{2}/2$~mm was chosen for future
tracking, as it still results in a sufficiently short simulation time.

With this, the verification of the tracking and interpolation performance of the
simulation is completed. We have proceeded to verify the ability of the FEM
software to generate field maps that agree with the exact solutions for
idealised coil and electrode shapes and the descriptive power of the derived analytical
equations of spin motion.

\subsection{Verification of analytical equations}\label{sec:analytical_verification}
A model of the experimental setup was created in \texttt{ANSYS Maxwell3D} with a
solenoid consisting of one main coil and two shimming coils on each side. The
coil parameters (current density, radius, position, length, etc.) were obtained
from a best fit of a measured field map of the superconducting
solenoid that will be used in the Phase~I muon EDM experiment. The circular coil
that produces the weakly-focusing field is positioned in the centre of the
solenoid. It produces a field that in the centre of the coil points in the opposite direction to that of the main
solenoid, which serves as a potential well in which muons are stored. The
simulation also includes a coaxial cylindrical electrode system
(\SI{20}{mm} inner and \SI{40}{mm} outer electrode radii) that provides the radial
electric field for the frozen-spin condition.
The FEM software is then used to calculate the field produced by the coils and
electrodes and to output it on a regular grid.

\begin{figure}[]
	\centering
	\includegraphics[width=\columnwidth]{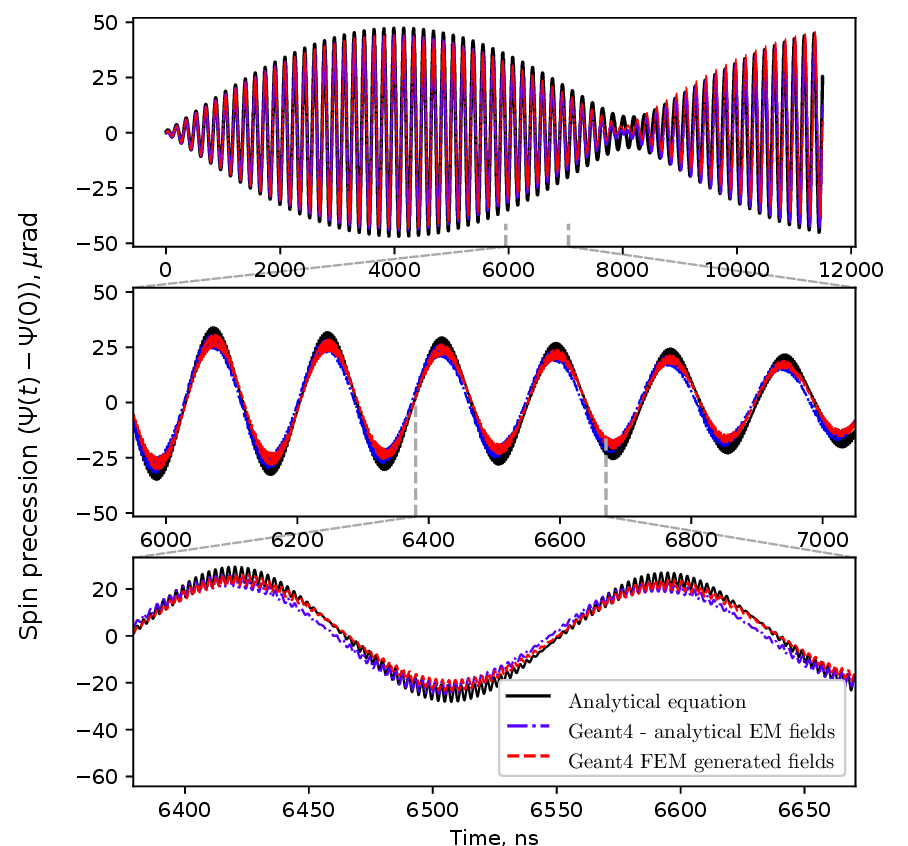}
	\caption{\raggedright Comparison between the analytical Eq.
		\eqref{eq:precession_theta_integral} and the \texttt{Geant4} spin tracking
		simulation.
		The initial parameters are arbitrarily set to $\vec S_0 =
			(-0.89, -0.46, 0.00)$, $\vec p_0 = (0.84, 0.55, 0.00) \times \SI{26.8}{\MeVc}$ and
		$\vec r_0 = (-16.95, 22.10, 5.00)\,$\SI{}{mm}. The muon momentum is intentionally not set ideally to the ideal muon momentum for the frozen-spin field of $E_\mathrm{f} = \SI{280}{kV/m}$ in order to highlight the $\omega_z$ oscillations of the spin. The three plots depict different timescales of the spin motion and show uncompensated $(g-2)$ precession (top), betatron oscillations (middle), and cyclotron oscillations (bottom).}
	\label{fig:good_agreement_ana_g4}
\end{figure}

A comparison between the analytical equations derived and the \texttt{Geant4}
spin tracking is shown in Fig.~\ref{fig:good_agreement_ana_g4}. The fields used
were either provided by exact analytical solutions for the weakly-focusing coil
and coaxial electrodes or field maps from numerical simulations. The initial
coordinates of the muon at the moment of storage were arbitrarily chosen (values
specified in the caption of Fig. \ref{fig:good_agreement_ana_g4}) for
illustration purposes. The electric field was set at such a value as to have
imperfect cancellation of the $(g-2)$ precession. Both the inner and outer
electrodes that generate the coaxial $E$-field are tilted around the same pivot
with respect to the $z$-axis by $\delta =\SI{0.01}{\degree}$ to highlight the
cyclotron oscillations. Simulations were performed for the \SI{1000}{m} track
length, or about 7 muon lifetimes.

The comparison shows very good agreement between the analytical equations and
the \texttt{Geant4} spin tracking performed using field maps or exact solutions. The weakly-focusing field gradient $\partial_z B_\rho$ (needed to calculate the field index $n$ in Eq. \eqref{eq:precession_theta_integral}) was calculated using an exact solution of a perfect circular current loop~\cite{NASAloop} at position $\rho = \rho_0$ and $z = 0$. This description of the
weakly-focusing field provides good estimates of the field strength and the betatron frequency. The difference between the exact and numerical approaches for field generation is negligible throughout the whole simulation time, demonstrating the good agreement between the two, also
when using realistic field maps from a physical coil (\SI{62}{mm} radius and \qtyproduct{5 x 5}{\mm} square cross section) generated by the finite-element method.

\subsection{Verification of geometric-phase
	equations}\label{sec:geometric_verification}
For the verification of the equations derived for the calculation of the
geometric phase accumulation, we consider two different cases: resonant and
non-resonant oscillations.
\subsubsection{Resonant oscillations}
Two periodic oscillations along perpendicular axes with equal angular
frequencies can occur due to the cyclotron motion of muons inside the $E$-field
used for the frozen-spin technique.
Consider the case described in Eq.~\eqref{eq:e_field_ref_frame}, where there is an angle between the
common axes of the electrodes and the solenoid magnetic
field, and the centre of the muon orbit is displaced by $\vec r_0$ with respect
to the centre of the electric field.
Equation~\eqref{eq:rotation_orbit_doesnt_matter}
shows that in this case, the net $(g-2)$ precession over a turn would be zero.
The net precession due to the longitudinal $E$-field component seen in the
muon reference frame would also be zero. However, the spin
will cause small oscillations around the longitudinal and radial axes with the
cyclotron angular frequency $\omega_\mathrm{c}$. A geometric phase will be observed if
the displacement of the orbit $\vec r_0$ is such that both oscillations are out
of phase.

To calculate the resulting geometric phase accumulation, we use
Eq.~\eqref{eq:berry_phase_equal_phases}, where $\omega = \omega_\mathrm{c}$. The
amplitudes $A_z$ and $A_\rho$ are the maximum spin precession angular
velocity resulting from the oscillations in the $z$ and radial $E$-field
components. Using the Thomas-BMT equation and Eq.~\eqref{eq:e_field_rotated}
they can be approximated as
\begin{align}
	A_z    & = -\frac{ea}{mc} \left[ 1 - \frac{1}{a(\gamma^2 - 1)}\right] \beta_\theta \frac{\max(E_z) -
	\min(E_z)}{2} \text{ and }                                                                           \\
	A_\rho & = -\frac{ea}{mc} \left[ 1 - \frac{1}{a(\gamma^2 - 1)}\right] \beta_\theta
	\frac{\max(E_\rho) - \min(E_\rho)}{2}.
	\label{eq:berry_omega_xy_efield}
\end{align}

\begin{figure*}[t]
	\centering
	\hspace*{-10pt}
	\subfloat[][$E$-field]{
        \scalebox{0.535}{\includegraphics{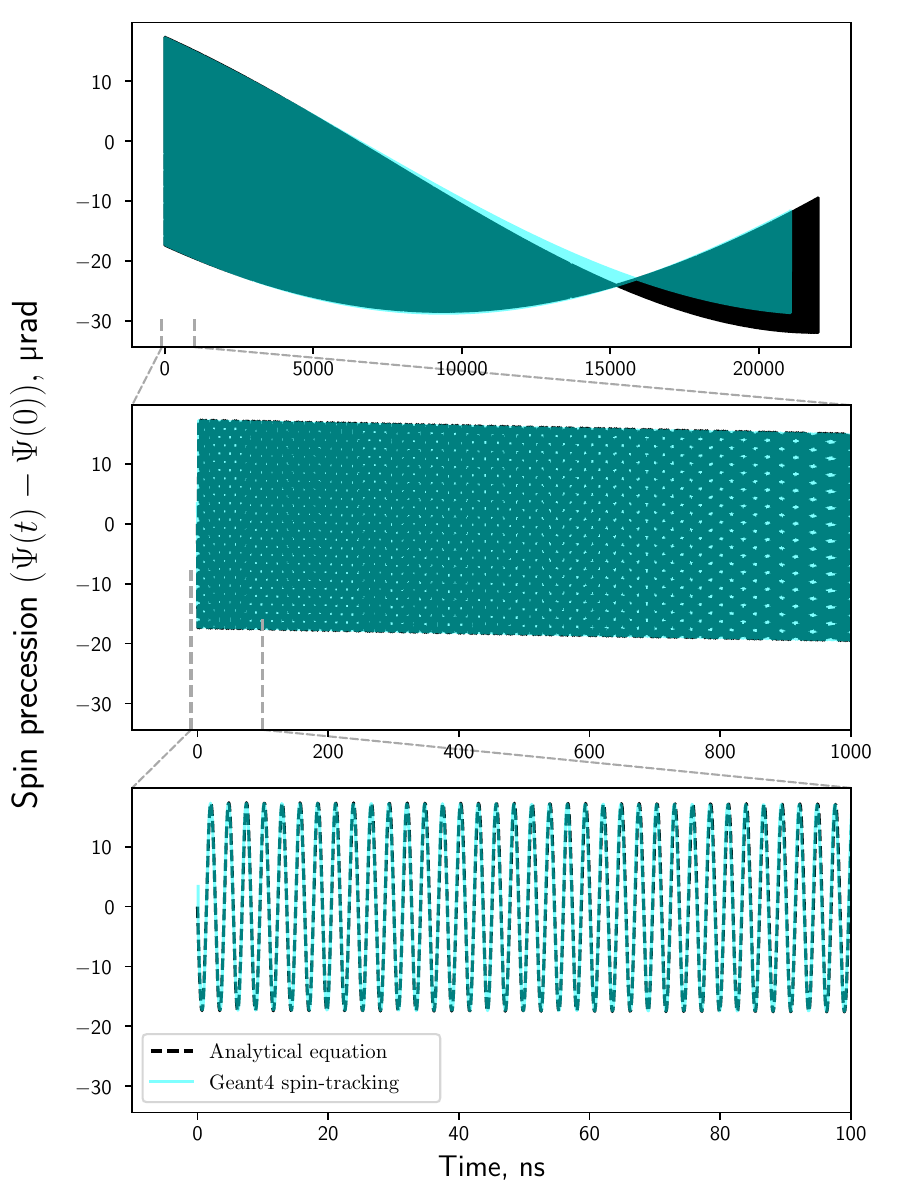}}
		\label{fig:berry_phases_efield}
	}
	\subfloat[][$B$-field]{
        \scalebox{0.535}{\includegraphics{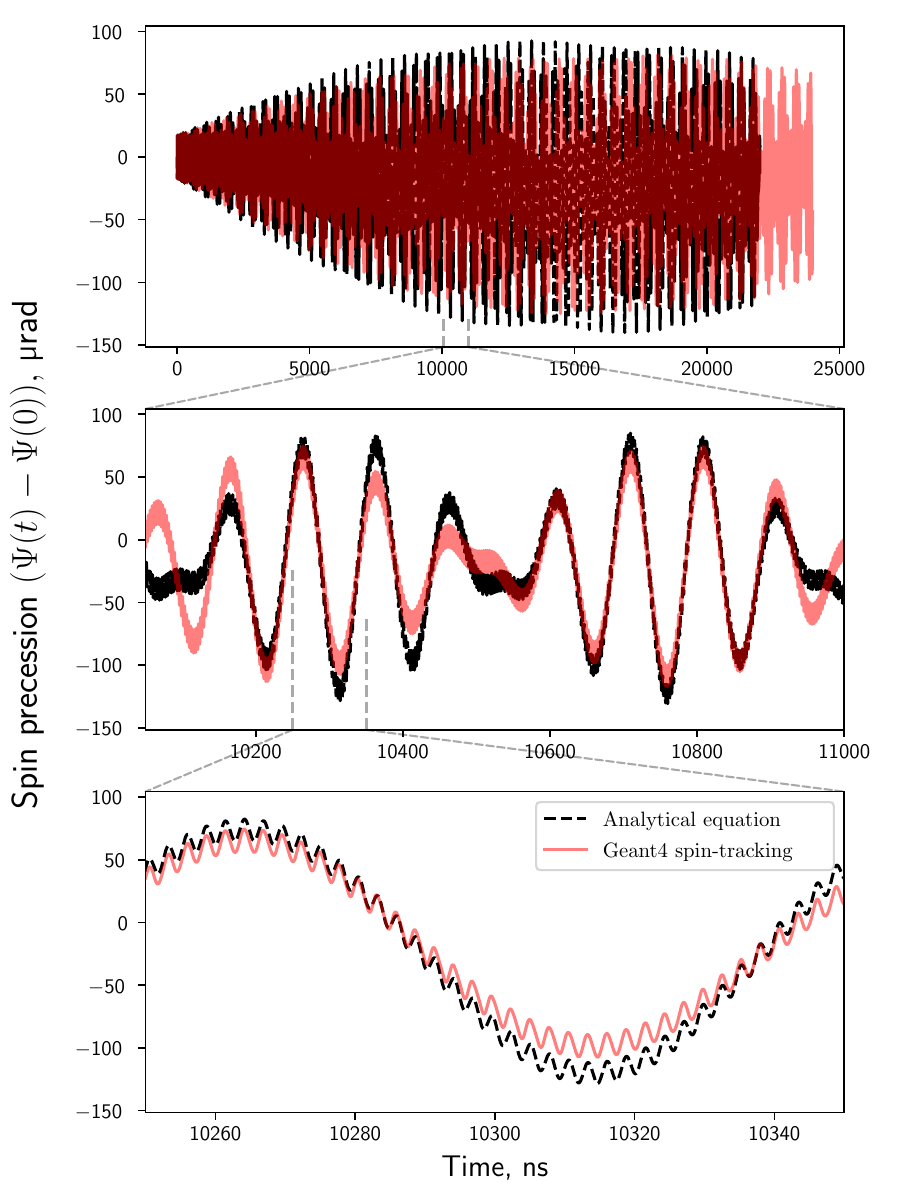}}
		\label{fig:berry_phases_bfield}
	}
	\caption[Accumulation of Berry's phase]{Accumulation of Berry's phase due to oscillations of the spin around
		perpendicular axes. Figure~\subref{fig:berry_phases_efield} shows the
		case of equal oscillation frequencies (resonant oscillations) around the two axes caused by changing $E$-field in the muon reference frame.
		Figure~\subref{fig:berry_phases_bfield} shows the Berry phase for the
		superposition of $B$-field oscillations with different frequencies (non-resonant oscillations) coupling to the $E$-field oscillations due to the cyclotron motion of the muon.}
	\label{fig:berry_phases_sim}
\end{figure*}

The \texttt{Geant4} simulation of the
experimental setup was prepared with exaggerated parameters to highlight the
phase accumulation effect to verify \eqref{eq:berry_phase_equal_phases}. The coaxial electrodes are tilted at \SI{15}{mrad}
and the centre of the muon orbit is displaced by \SI{5}{mm}. The radial
$E$-field is lower than that required for the frozen-spin condition to allow for
a residual $(g-2)$ precession. The phase between the two oscillations is set to
the worst-case scenario, i.e. $\pi/2$. The comparison between the equations and
the spin tracking simulation is shown in Fig.~\ref{fig:berry_phases_efield}. For
the purpose of comparison, we assume classical parallel transport, since the
simulation software is not capable of simulating quantum behaviour. In reality,
phase accumulation will be half of the value obtained.

\subsubsection{Non-resonant oscillations}
To validate Eq.~\eqref{eq:berry_phase}, the same simulation conditions
as the previous comparison were used, but with an additional oscillating radial
$B$-field. Spin oscillations caused by the $B$-field coupled with those in $E_z$ that are due to the \SI{8.7}{mrad} tilt of the electrode system. The source of the radial field in the simulation is the anti-Helmholtz pair of coils, which is a realistic scenario as in practice such radial oscillations can be induced by a residual
$B$-field from the magnetic kick. The $B$-field in the comparison is generated by superposition of two currents of the form $I(t) = I_0 \sin(\omega t + b_0)$ with different amplitudes and initial phases and angular velocities at $\pm 5$\% of the cyclotron angular velocity (\SI{2.3}{rad/ns}). The amplitudes of the two sinusoidal components are \SI{10}{A} and \SI{5}{A}, which is more than an order of magnitude higher than the expected residual current after the magnetic kick. This is done to highlight the geometric phase effect, as otherwise it would be negligible. The effect is also enhanced by setting the tilt of the electrode system to such a large value, again for the purpose of illustrating the geometric phase effect.

The results
of the comparison are presented in Fig.~\ref{fig:berry_phases_bfield}. A very
good agreement between theoretical prediction and spin-tracking simulations is observed
on the microsecond scale ($g-2$ precession) and the nanosecond scale (cyclotron
oscillations) over about 10 muon lifetimes. The overall behaviour of the geometric phase is captured by the
analytical equations, though there are small differences between the predicted and
observed beating patterns in case of the $B$-field coupling. The oscillation amplitude of the radial $B$-field generated from the magnetic kick was calculated at the $(x, y, z) = (\rho_0, 0, 0)$ position, whereas the muon experiences longitudinal and horizontal betatron oscillations, thus resulting in a slightly different $B$-field. Despite this approximation, the agreement between the simple calculation and the detailed spin tracking is satisfactory. Other sources of difference between the analytical and simlation approaches could come from the discretisation of the $B$-field
used in the numerical simulation, which introduces additional
high-frequency noise, and from the calculation of the cyclotron frequency, which is proportional to be $B$-field. In the analytical calculation we have used the \SI{3}{T} main solenoid field, which does not include the effects of the weakly-focusing coil or the variable field due to the kick. Differences could also be due to the motion of the muons in space, since the generated field is the sum of two sines, but the field in the rest frame of the particle would have higher- and lower-frequency terms.

\section{Striped electrode system}\label{sec:striped_electrodes}
The limits shown in section~\ref{sec:e-field_uniformity} place stringent
constraints on the alignment and shape of the electrode system. Although the
simplest way to achieve a radial $E$-field is to employ a coaxial cylindrical
structure, these limits could be achieved more easily with a more complex setup,
e.g. inner and outer electrode geometries that are composed of individual wires
instead of a uniform cylindrical foil.

\begin{figure}[htb]
	\centering
	\includegraphics[width=\columnwidth]{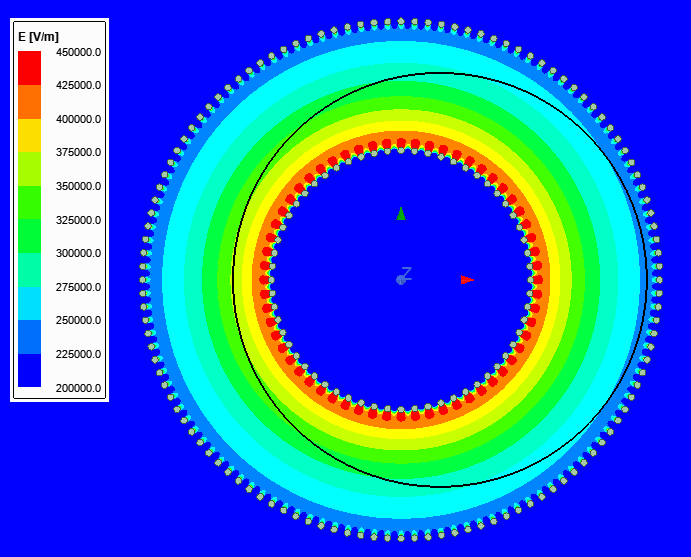}
	\caption{Radial electric field for \SI{20}{mm} inner charged electrode
		radius and \SI{40}{mm} outer grounded electrode radius when the electrodes
		are composed of \SI{1}{mm} diameter wires. The outer electrode has 120 and
		the inner 60 wires. The inner electrode high-voltage is \SI{-6200}{kV}. The
		black circle shows a \SI{6}{mm} displaced orbit.}
	\label{fig:ansys_striped_field}
\end{figure}

Such a setup would introduce radial non-uniformity along the circular muon
orbit, which could lead to the accumulation of a geometric phase. However, as
seen in Fig.~\ref{fig:berry_b_limit}, the limits on the electric field
uniformity are quite relaxed for high-frequency oscillations. For example,
separating the electrodes into 100 longitudinal strands would create radial
field non-uniformity. If there is a tilt of the electrode system, that would
translate to an $E_z$ non-uniformity as well. However, the frequency of the
$E$-field oscillations due to this non-uniformity will be at about \SI{40}{GHz}.
At such high frequency oscillations of the radial component on the level of even
50\% would produce a negligible geometrical phase accumulation. Note, however,
that the geometrical phase calculations assume adiabatic processes and such
extreme conditions might violate that assumption. FEM simulations of the
electric field generated by an electrode segmented into segmented into 60 parts, each consisting of a \SI{1}{mm} diameter wire and a grounded electrode segmented into 120 wires show that the
amplitude of the high-frequency field oscillations are less than 1\% even for a
displacement of \SI{6}{mm} of the centre of the muon orbit. The transverse cross
section of the electric field and the displaced orbit are shown in
Fig.~\ref{fig:ansys_striped_field}. The geometrical phase effect of such an
electrode structure is shown by the red ellipsoid in
Fig.~\ref{fig:berry_e_limit}.

The main reason to prefer a wired electrode setup is that in practice it would
be easier to construct uniform electrodes compared to using cylindrical foils.
One can devise a setup where the wires are connected to piezoelectric actuators
that allow for very fine control of the position. The actual position of the
wire can then be measured with sub-micrometer precision using optical or
capacitive distance sensors. The produced electric field can then be precisely
simulated using finite-element methods.

Another advantage of a striped system is the reduced material budget, which
would increase the path length of decay positrons and reduce their scattering.
Compared to a cylindrical foil, it would also lead to significantly less eddy
currents from the magnetic kick.

\end{document}